\documentclass[letter]{jpsj3}
\usepackage{txfonts}
\usepackage{multirow}

\title{Holonomy-based Diagnostic of Strain Compatibility in Birefringence Imaging of Stress-induced Ferroelectric SrTiO$_3$}

\author{Hirotaka Manaka$^1$\thanks{E-mail address: manaka@eee.kagoshima-u.ac.jp}, Kazuma Seike$^1$, and Yoko Miura$^2$}

\inst{$^1$Graduate School of Science and Engineering, Kagoshima University, Korimoto, Kagoshima 890-0065, Japan\\
$^2$National Institute of Technology, Suzuka College, Shiroko-cho, Suzuka, Mie 510-0294, Japan }

\abst{We introduce a holonomy-based geometric diagnostic for birefringence-derived director fields and apply it to stress-induced ferroelectric SrTiO$_3$. Treating the director as a line field in $\mathbb{R}P^2$, we define a holonomy angle $\omega$ from residual rotations accumulated along closed loops in real space and compare it with a conventional local-gradient metric. Whereas the gradient quantifies local orientational variation, $\omega$ probes the global compatibility of rotations along closed paths. The resulting $\omega$ map cannot be reproduced by simple coarse-graining of local gradients, indicating sensitivity to loop-level orientational incompatibility. Analysis of alignment of holonomy rotation axes reveals a cooling-induced reorganization of the electromechanical response, consistent with strain- or stress-related inhomogeneity above the ferroelectric transition and additional ordering below it. These results demonstrate holonomy as a loop-based geometric diagnostic of strain compatibility in orientational fields derived from birefringence.}

\begin{document}
\maketitle

Ferroic materials such as ferroelectrics and ferroelastics often develop spatially inhomogeneous textures of polarization and/or lattice strain, including domains, domain walls, junctions, and vortex-like structures, depending on temperature ($T$), stress, defects, and boundary conditions \cite{I1,I2,I3}. Because these textures are closely linked to phase-transition mechanisms and functional responses, their real-space characterization is essential. Optical birefringence imaging is particularly valuable, as it provides direct real-space information on strain/stress and orientational (director) structures via the polarization state of transmitted light \cite{azz,manaka-bf1,manaka-bf7}.

Conventional birefringence-image analyses typically rely on local gradient--based indicators, such as nearest-neighbor angular variation of the director field (``grad''), which capture local variations but cannot distinguish globally compatible fields from path-dependent ones. To address this limitation, we introduce the holonomy angle $\omega$, defined as the residual rotation accumulated along a closed loop \cite{h1,ho1}, for the reconstructed $\mathbb{R}P^2$ director field rather than for the optical wave itself. For an integrable orientational field, this residual vanishes even in the presence of large local gradients, whereas a finite $\omega$ indicates path-dependent non-integrability. Thus, $\omega$ provides a loop-based measure of orientational incompatibility beyond local-gradient metrics. Here, we apply this framework to experimentally derived birefringence-based $\mathbb{R}P^2$ fields and demonstrate its capability to detect loop-level incompatibility not captured by conventional approaches.

We revisit previously reported birefringence-imaging data for stress-induced ferroelectric SrTiO$_3$ \cite{manaka-bf8,manaka-bf10}. Under an applied stress of 231~MPa along [001] during cooling, the sample undergoes a cubic-to-tetragonal structural transition at $T_{\rm c}=114$--120~K, followed by a ferroelectric transition at $T_{\rm F}=19$--30~K \cite{manaka-bf8,manaka-bf12,manaka-bf13}. Prior work reconstructed a spatial map (302$\times$140 pixels) of $T_{\rm F}$ [Fig.~1(a)] and showed that higher-$T_{\rm F}$ regions correlate with stress concentration \cite{manaka-bf13}. To relate this inhomogeneity to local electromechanical response, the stress-induced polarization is expressed to first order as $P_i \approx e_{ijk}\,\varepsilon_{jk} + \mu_{ijkl}\,\partial_l\varepsilon_{jk}$, where the first and second terms denote piezoelectric and flexoelectric contributions, respectively \cite{Fl3}. Since $\rho_{\mathrm b}=-\nabla\cdot\boldsymbol{P}$, regions with $\nabla\cdot\boldsymbol{P}\neq 0$ correspond to local bound-charge accumulation and enhanced electromechanical inhomogeneity. A prior statistical causal analysis further showed that $T_{\rm F}$ is elevated in stress-concentrated regions, with the dominant contribution attributed to flexoelectricity \cite{manaka-bf14}. In this context, correlations with $T_{\rm F}$ reflect underlying electromechanical inhomogeneity.

In continuum elasticity, compatible strain fields satisfy the Saint--Venant compatibility condition, whereas incompatibility arises from defects or plastic deformation \cite{Sa1,Sa2}. Because birefringence reflects strain-induced anisotropy, the reconstructed director field represents local strain orientation. In this context, holonomy provides a geometric test of strain compatibility: vanishing $\omega$ indicates an integrable field, whereas finite $\omega$ signifies path-dependent incompatibility and is therefore associated with electromechanical inhomogeneity arising from spatial strain variations. Here, we do not claim direct observation of flexoelectric polarization; rather, we test whether such non-integrable optical textures preferentially occur in regions where electromechanical incompatibility is expected. Accordingly, $\omega$ is interpreted as a geometric signature of inconsistency in the strain-induced orientational field, not as direct evidence of flexoelectric polarization or topological non-triviality.

For the holonomy analysis, we use 575-nm birefringence data \cite{manaka-bf8}. At each pixel $(x,y)$ and $T$, the polarization state is represented by a normalized Stokes vector, from which the director $\boldsymbol{n}^{(T)}(x,y)$ is defined on the Poincar\'e sphere \cite{Supply}. Because birefringence imaging identifies $\boldsymbol{n}^{(T)}\sim -\boldsymbol{n}^{(T)}$, the director is treated as a line field in $\mathbb{R}P^2$. Finite $\omega$ can arise from geometric (non-topological) incompatibility even in topologically trivial fields and does not necessarily imply topological defects \cite{h1,ho1,ho2,h2}. In the present data, the line field is topologically trivial; therefore, nearest-neighbor minimal rotations can be defined without introducing branch cuts \cite{Supply}.

The shortest rotation between neighboring directors is represented by a unit quaternion ${q_e}^{(T)}\in\mathrm{SU}(2)$ associated with an oriented nearest-neighbor edge $e$, with ${q_{-e}}^{(T)}=({q_e^{(T)}})^{-1}$ for the reversed edge. This representation provides a compact, singularity-free description of three-dimensional rotations, enabling composition by simple multiplication \cite{Supply}. We compose these local rotations along the boundary $\partial\Box_L(x,y)$ of a $L\times L$-pixel square loop:

\begin{equation}
Q_L^{(T)}(x,y)= \prod_{e\in\overrightarrow{\partial\Box_L(x,y)}} q_e^{(T)} \in \mathrm{SU}(2),
\end{equation}

\noindent where the product follows the loop traversal order \cite{Supply}. Although Eq. (1) is written for an $L\times L$ square loop in the experimentally available $xy$ image plane, the construction itself is not restricted to two-dimensional real space. Since the essential operation is the assignment of a minimal rotation $q_e$ to each oriented edge, the same procedure can in principle be applied to a three-dimensional director field by multiplying the corresponding edge rotations along closed paths in $xyz$ space. Writing $Q_L^{(T)}(x,y)=(w_L^{(T)}(x,y),\boldsymbol{v}_L^{(T)}(x,y))$ with the scalar part chosen as $w_L^{(T)}(x,y)\ge0$, we define the holonomy angle:

\begin{equation}
\omega_L^{(T)}(x,y) :=2\arctan2\bigl(\|\boldsymbol{v}_L^{(T)}(x,y)\|,\,w_L^{(T)}(x,y)\bigr)\in[0,\pi], 
\end{equation}

\noindent where $\boldsymbol{v}_L^{(T)}(x,y)$ is the vector part of the quaternion, whose direction defines the unit rotation axis $\hat{\boldsymbol{u}}_L^{(T)}(x,y)=\boldsymbol{v}_L^{(T)}(x,y)/\|\boldsymbol{v}_L^{(T)}(x,y)\|$. Thus, $\omega_L^{(T)}(x,y)=0$ corresponds to a vanishing loop residual, whereas larger values indicate stronger loop-level, path-dependent inconsistency in the locally composed director rotations.

Motivated by the previously reported $\sim$10-pixel scale of slip-plane and ferroelectric textures, we adopt $L=10$ (a $10\times10$-pixel loop) as the nominal analysis scale \cite{manaka-bf12,manaka-bf13,manaka-bf15}. Preliminary calculations for $L=2$--20 show that the qualitative spatial patterns and $T$ dependence remain unchanged, indicating robustness to the choice of scale \cite{Supply}. Near the left and top boundaries in the same image frame as Fig.~1(a), defined with the origin at the lower-right corner, an $L=10$ loop extends beyond the field of view. In these regions, the largest admissible square loop is used. These regions are retained for visualization only and excluded from statistical aggregation.

Birefringence images were acquired during continuous cooling from 300.0 to 14.1~K. To visualize spatial patterns, the data are grouped into 5-K-wide $T$ intervals, with $\mathcal{T}_{5W}$ denoting the set of $T$ values in each interval \cite{Supply}. For each window, we define the representative $\omega$ map at scale $L$ as

\begin{equation}
\overline{\omega}_L^{(5W)}(x,y) := \operatorname{median}_{T\in\mathcal{T}_{5W}} \omega_L^{(T)}(x,y).
\label{eq:S1_rep_omega}
\end{equation}

\noindent Figure~1(b) shows the $\overline{\omega}_{10}^{(5W)}$ map for $(40.0~{\rm K},45.0~{\rm K}]$ (42K-wnd), where ``wnd'' denotes a temperature window. Maps for $[14.1~{\rm K},15.0~{\rm K}]$ (14K-wnd) and $(130.0~{\rm K},135.0~{\rm K}]$ (132K-wnd) are also provided in Ref. \cite{Supply}, where the effect of the reduced $|\mathcal{T}_{5W}|$ in the 14K-wnd is examined. The 42K-wnd is of interest because the inferred $T_{\rm F}$ is typically below 30~K, with a maximum of 31.31~K (Fig.~1(a)) \cite{manaka-bf13}. Thus, it satisfies $T>T_{\rm F}$ for all pixels while remaining close to the ferroelectric regime, providing a reference for examining the relation between loop-level optical non-integrability and strain-gradient-induced electromechanical inhomogeneity in the paraelectric state \cite{Fl2,Fl1}. To identify spatially localized anomalies in the holonomy field, we extract local maxima of the representative map using non-maximum suppression with a minimum separation of 5 pixels \cite{Supply,NMS}. The circles in Fig.~1(b) mark the detected local maxima; their size is chosen to be smaller than the minimum separation to avoid overlap and indicate peak centers.

For comparison, Fig.~1(c) shows the representative grad map $\overline{\rm grad}^{(5W)}(x,y)$ for the 42K-wnd, constructed using the same pixel-wise median procedure as in Eq. (\ref{eq:S1_rep_omega}) \cite{Supply}. Here, grad denotes the nearest-neighbor angular variation of the director field on $\mathbb{R}P^2$, defined as the mean magnitude of sign-invariant orientation differences between neighboring pixels. The comparison with the inferred $T_{\rm F}$ map is relevant because Fig.~1(a) highlights regions with enhanced polarization-related responses and stress concentration. Both the $\overline{\omega}_{10}^{(5W)}$ and $\overline{\rm grad}^{(5W)}$ maps tend to increase in regions with higher $T_{\rm F}$, although their spatial patterns differ. Because $\overline{\omega}_{10}^{(5W)}$ represents rotation accumulated over a finite $L=10$ loop, it exhibits a smoother spatial profile than $\overline{\rm grad}^{(5W)}$. In particular, the elongated stripe-like structure extending toward the top edge appears prominently in the $\overline{\rm grad}^{(5W)}$ map but fragments into localized clusters in the $\overline{\omega}_{10}^{(5W)}$ map. Consequently, large $\overline{\omega}_{10}^{(5W)}$ values typically occur within broader high-grad regions but identify locations where the orientational field is geometrically incompatible, rather than simply exhibiting large gradient magnitude.

To quantify spatial similarity, we compare high-value regions across the maps using the intersection over union (IoU) and Dice coefficient, which measure overlap between binary masks \cite{Supply}. Defining masks by the upper 10\% of pixels in each map---motivated by clustering analyses indicating a $\sim$10\% area fraction of stress-concentrated regions \cite{manaka-bf12}---we obtain (${\rm IoU}_{10\%}$, ${\rm Dice}_{10\%}$) = (0.420, 0.591) for Figs.~1(a) and~1(b), (0.455, 0.625) for Figs.~1(a) and~1(c), and (0.391, 0.562) for Figs.~1(b) and~1(c). These trends persist under variations in analysis parameters, including loop size ($L=5$ and 20), gradient smoothing, and percentile thresholds \cite{Supply}. Robustness is further supported by spatial statistical tests. A density scatter plot of grad versus $\omega$ shows moderate correlation with substantial spread in $\omega$ even within high-grad regions. Large $\omega$ values typically occur as localized subsets within broader high-grad regions, whereas extended grad features often lack large $\omega$. This relation indicates that the holonomy angle isolates localized, loop-level incompatibilities embedded within more extended gradient-dominated patterns, rather than serving as a smoothed measure of gradient magnitude, consistent with its intrinsically nonlocal, loop-based nature.

To examine the $T$ dependence of $\omega$, we evaluate spatial summaries at each $T$. Because $\omega_L^{(T)}(x,y)$ is highly inhomogeneous and skewed---most pixels are near zero, with only a small fraction taking large values---we separate the global background from the high-$\omega$ subset by defining

\begin{equation}
\omega_{L, \alpha}^{(T)} := \operatorname{median}_{(x,y)\in\Omega_{L, \alpha}^{(T)}} \omega_L^{(T)}(x,y)
\qquad
\alpha\in\{{\rm all},10\%\},
\end{equation}

\noindent where $\Omega_{L, {\rm all}}^{(T)}$ denotes all valid pixels for an $L\times L$ loop, and $\Omega_{L, 10\%}^{(T)}$ the top $10\%$ of pixels ranked by $\omega_L^{(T)}(x,y)$ for the same $L$. The spatial median reduces sensitivity to extreme values. The data are then rebinned into 1-K-wide intervals. Denoting by $\mathcal{T}_{1W}$ the set of $T$ values within each interval, we define

\begin{equation}
\overline{\omega}_{L, \alpha}^{(1W)}
=
\frac{1}{|\mathcal{T}_{1W}|}
\sum_{T\in\mathcal{T}_{1W}}
\omega_{L, \alpha}^{(T)} \qquad
\alpha\in\{{\rm all},10\%\}.
\end{equation}

\noindent Figure~2(a) shows $\overline{\omega}_{10, {\rm all}}^{(1W)}$ and $\overline{\omega}_{10, 10\%}^{(1W)}$. Their magnitudes differ by more than an order of magnitude: most pixels have $\omega_L^{(T)}(x,y)\simeq 0$, whereas large values are confined to limited regions. Nevertheless, both curves exhibit clear anomalies near $T_{\rm c}$ and $T_{\rm F}$, with similar behavior observed for other thresholds \cite{Supply}. These results indicate that statistically significant holonomy arises primarily in a restricted subset of pixels and evolves characteristically, in a path-dependent manner, across both transitions. In this sense, $\overline{\omega}_{10, 10\%}^{(1W)}$ serves as a sensitive indicator of the evolution of loop-level orientational incompatibility in the optical orientational field.

We next examine the spatial organization of the rotation axis associated with the loop quaternion. While $\omega_L^{(T)}(x,y)$ quantifies the magnitude of the loop residual, it does not indicate whether the corresponding rotation axes are aligned or broadly distributed on the Poincar\'e sphere. Because the optical director is closely related to the birefringence fast-axis orientation, non-integrability may arise from spatially incompatible strain or strain-gradient components in the underlying electromechanical response. To quantify this aspect, we introduce the sign-invariant second-rank orientation tensor:

\begin{equation}
\boldsymbol{A}_{L, \alpha}^{(T)}
=
\frac{1}{|\Omega_{L, \alpha}^{(T)}|}
\sum_{(x,y)\in\Omega_{L, \alpha}^{(T)}}
\hat{\boldsymbol{u}}_L^{(T)}(x,y)\,
\hat{\boldsymbol{u}}_L^{(T)}(x,y)^{\mathsf T},
\end{equation}

\noindent where $\alpha\in\{{\rm all},10\%\}$ denotes either all pixels or the subset comprising the upper 10\% ranked by $\omega_L^{(T)}(x,y)$ for an $L\times L$ loop. Because $\hat{\boldsymbol{u}}_L^{(T)}(x,y)\sim-\hat{\boldsymbol{u}}_L^{(T)}(x,y)$, a second-moment tensor is used instead of a first-moment average. Let $\lambda_{L, \alpha}^{\rm max}(T)$ be the largest eigenvalue of $\boldsymbol{A}_{L, \alpha}^{(T)}$, and define

\begin{equation}
S_{L, \alpha}^{(T)} := \frac{3\lambda_{L, \alpha}^{\rm max}(T)-1}{2},
\end{equation}

\noindent where $S_{L, \alpha}^{(T)}\simeq1$ indicates strong axis alignment, whereas $S_{L, \alpha}^{(T)}\simeq0$ corresponds to an approximately isotropic axis distribution \cite{S1,S2}. The spatial alignment of the holonomy rotation axes reflects, in a geometric sense, the organization of the birefringence director field and is consistent with the dominant fast-axis orientation \cite{Supply}.

Figure~2(b) shows $\overline{S}_{10, \alpha}^{(1W)}$, obtained by averaging $S_{10, \alpha}^{(T)}$ over each 1-K-wide interval, as in Eq.~(5). Although $\overline{S}_{10, {\rm all}}^{(1W)}$ and $\overline{S}_{10, 10\%}^{(1W)}$ exhibit similar trends, $\overline{S}_{10, 10\%}^{(1W)}$ is systematically smaller, indicating that high-$\omega$ regions exhibit greater axis disorder. Around 19--30~K, spatial variation in local $T_{\rm F}$ is associated with enhanced inhomogeneity and reduced $\overline{S}_{10, \alpha}^{(1W)}$. For $T<T_{\rm F}$, the increase in $\overline{S}_{10, \alpha}^{(1W)}$ is consistent with partial alignment of $\hat{\boldsymbol{u}}_{10}^{(T)}(x,y)$ within ferroelectric domains; however, the order does not recover to the high-$T$ level, likely because the sample does not reach a single-domain state. Taken together, Figs.~2(a) and~2(b) indicate that holonomy analysis captures anomalies near both $T_{\rm c}$ and $T_{\rm F}$ at complementary levels: $\overline{\omega}_{10, \alpha}^{(1W)}$ quantifies the strength of loop-level non-integrability, whereas $\overline{S}_{10, \alpha}^{(1W)}$ characterizes the spatial organization of the associated rotation axes.

To examine how the organization of rotation axes evolves with decreasing $T$, we analyze the axis-order field in real space. The local axis-order parameter $S_L^{(T)}(x,y)$ at a given $T$ provides a spatial map of rotation-axis alignment. Here, in contrast to $S_{L,\alpha}^{(T)}$, which is defined over subsets of pixels, $S_L^{(T)}(x,y)$ denotes the local quantity evaluated at each pixel. To improve statistical robustness, the analysis is performed over 5-K-wide intervals ($\mathcal{T}_{5W}$). For each window, we define the representative axis-order map:

\begin{equation}
\overline{S}_L^{(5W)}(x,y)
=
\frac{1}{|\mathcal{T}_{5W}|}
\sum_{T\in\mathcal{T}_{5W}} S_L^{(T)}(x,y),
\end{equation}

\noindent and its change relative to the high-$T$ reference window (132K-wnd), which lies above $T_{\rm c}$:

\begin{equation}
\Delta S_L^{(5W)}(x,y)
=
\overline{S}_L^{(132\mathrm{K\mbox{-}wnd})}(x,y)
-
\overline{S}_L^{(5W)}(x,y).
\end{equation}

\noindent Thus, $\Delta S_L^{(5W)}(x,y)$ highlights regions where the spatial organization of loop-rotation axes evolves during cooling. To ensure consistent comparison across windows, $\Delta S_L^{(5W)}$ is evaluated on a common set of reliable pixels defined by a global quality filter based on a global threshold applied to $\overline{\omega}_L^{(5W)}(x,y)$, excluding pixels with small values \cite{Supply}. This representation highlights, in a geometric sense, cooling-induced deviations of the axis-order structure from the high-$T$ configuration. Excluded pixels are shown transparently.

Figure~3 shows $\Delta S_{10}^{(5W)}$ maps for the 14K-wnd and 42K-wnd; maps for other windows are also provided in Ref. \cite{Supply}. These maps reveal pronounced real-space reorganization of axis ordering during cooling. Although the maximum values of $|\Delta S_{10}^{(5W)}|$ are comparable for the 14K-wnd and 42K-wnd, their spatial patterns differ markedly. In the 42K-wnd, $\Delta S_{10}^{(5W)}$ exhibits stripe-like structures extending from the upper left to the lower right, with a spatial trend similar to the grad map in Fig.~1(c). In contrast, the 14K-wnd shows stripes with the opposite orientation, and its overall spatial organization differs substantially from both $\overline{\rm grad}^{(5W)}$ and $\overline{\omega}_{10}^{(5W)}$ maps. This contrast suggests that the 42K-wnd pattern is largely consistent with strain- or stress-related inhomogeneity, whereas the 14K-wnd pattern may reflect additional ordering processes emerging in the ferroelectric regime.

To quantify reorganization of the spatial pattern of $\Delta S_{10}^{(5W)}$ between neighboring $T$ windows, we compare adjacent windows using similarity metrics \cite{Supply}. Pronounced changes occur near $T_{\rm c}$ and $T_{\rm F}$, whereas the spatial pattern remains largely preserved elsewhere. A modest local enhancement is observed between the 42K-wnd and 37K-wnd, suggesting the onset of spatial rearrangement above $T_{\rm F}$. This behavior is consistent with the spatially distributed nature of the inferred $T_{\rm F}$ and the gradual development of electromechanical inhomogeneity associated with stress-induced ferroelectricity.

In summary, holonomy analysis of birefringence imaging provides information complementary to conventional grad-based measures. $\overline{\omega}_{10}^{(5W)}(x,y)$ quantifies the strength and spatial localization of loop-level non-integrability, whereas $\Delta S_{10}^{(5W)}(x,y)$ captures the $T$-dependent reorganization of the rotation-axis structure. Thus, holonomy serves as a loop-based geometric diagnostic of strain compatibility, capturing path-dependent orientational incompatibility beyond local-gradient measures in birefringence-derived fields. Applied to stress-induced ferroelectric SrTiO$_3$, it reveals cooling-induced reorganization of the electromechanical response and clarifies its spatial organization and $T$ evolution. More generally, the present framework is not limited to birefringence-based measurements. In principle, it can be applied to any system for which a spatially resolved orientational or vector field can be reconstructed.

\section*{Acknowledgments}

This work was partially supported by Grants-in-Aid for Scientific Research (KAKENHI; Grant Nos. JP23K03283 and JP25K08487) from the Japan Society for the Promotion of Science.

\clearpage

\begin{figure}[b]
\begin{center}
\includegraphics[width=10cm]{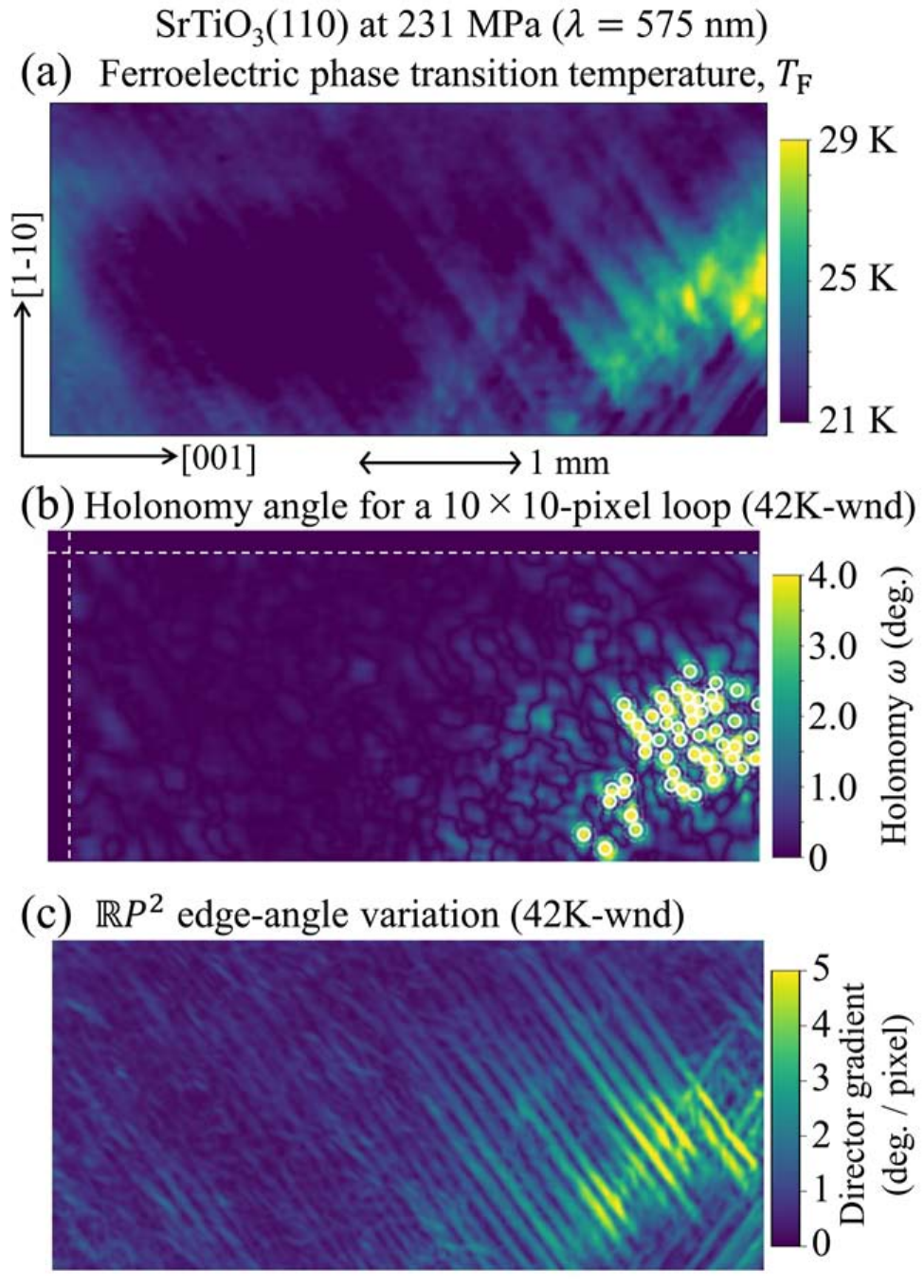}
\end{center}

\caption{(Color online) Real-space maps (302$\times$140 pixels) of (a) the ferroelectric transition temperature $T_{\rm F}$, (b) the holonomy angle $\overline{\omega}_{10}^{(5W)}(x,y)$ computed using $10\times10$-pixel loops, and (c) the nearest-neighbor $\mathbb{R}P^2$ edge-angle variation map $\overline{\rm grad}^{(5W)}(x,y)$, both evaluated over the 42K-wnd (40.0~K, 45.0~K]. In panel (b), regions above and to the left of the dashed lines are shown for visualization only, as the square loop extends beyond the field of view. The circles mark representative local maxima of $\overline{\omega}_{10}^{(5W)}$, identified via two-dimensional non-maximum suppression; their size is set smaller than the minimum separation scale to clearly indicate the peak centers.}
\end{figure}

\begin{figure}[b]
\begin{center}
\includegraphics[width=10cm]{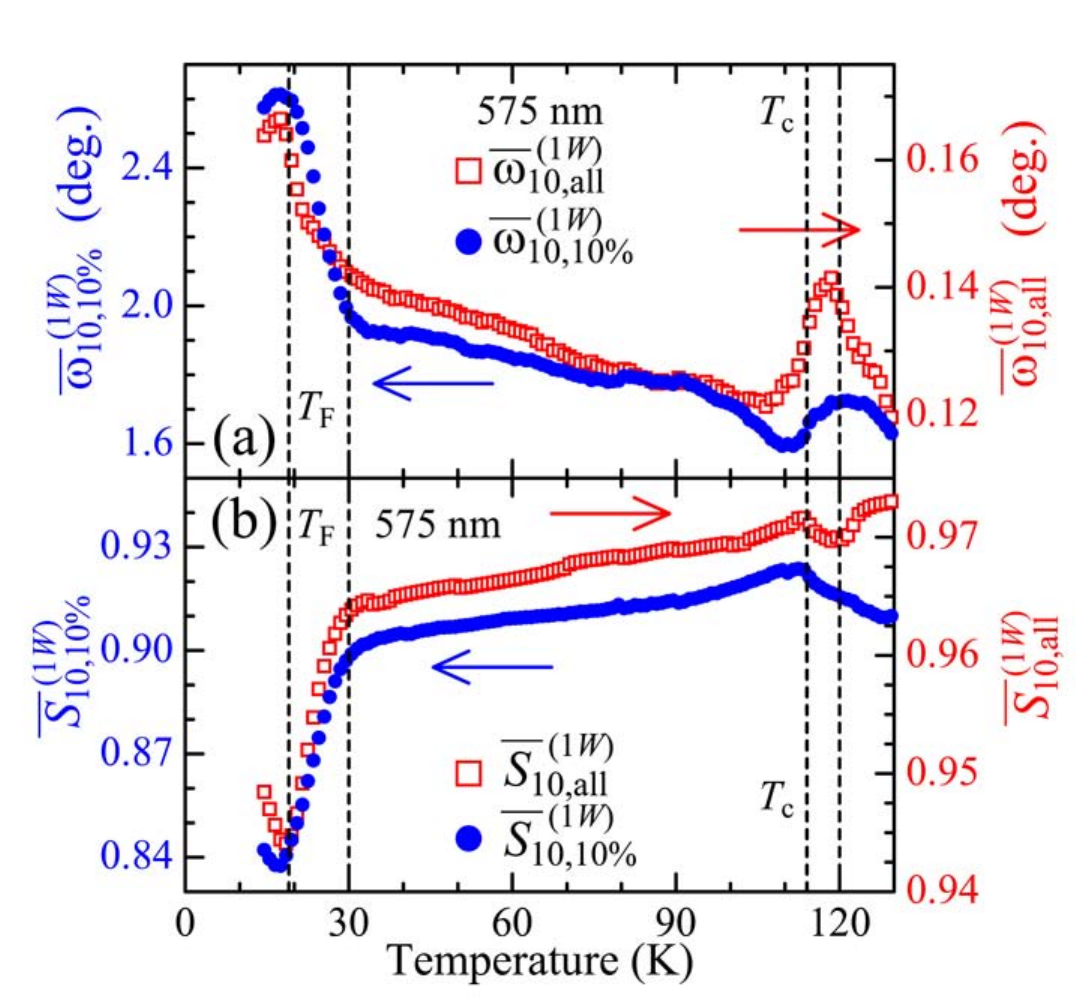}
\end{center}
\caption{(Color online) Temperature dependences of holonomy-related quantities for $L=10$, aggregated over 1-K-wide temperature windows. (a) Holonomy angle $\overline{\omega}_{10, \alpha}^{(1W)}$ and (b) axis-alignment order parameter $\overline{S}_{10, \alpha}^{(1W)}$. Here, $\alpha={\rm all}$ denotes aggregation over all valid pixels, whereas $\alpha=10\%$ denotes aggregation over the top 10\% of pixels ranked by $\omega_{10}^{(T)}(x,y)$. Dashed vertical lines indicate the reported ranges of $T_{\rm c}$ and $T_{\rm F}$.}
\end{figure}

\begin{figure}[b]
\begin{center}
\includegraphics[width=10cm]{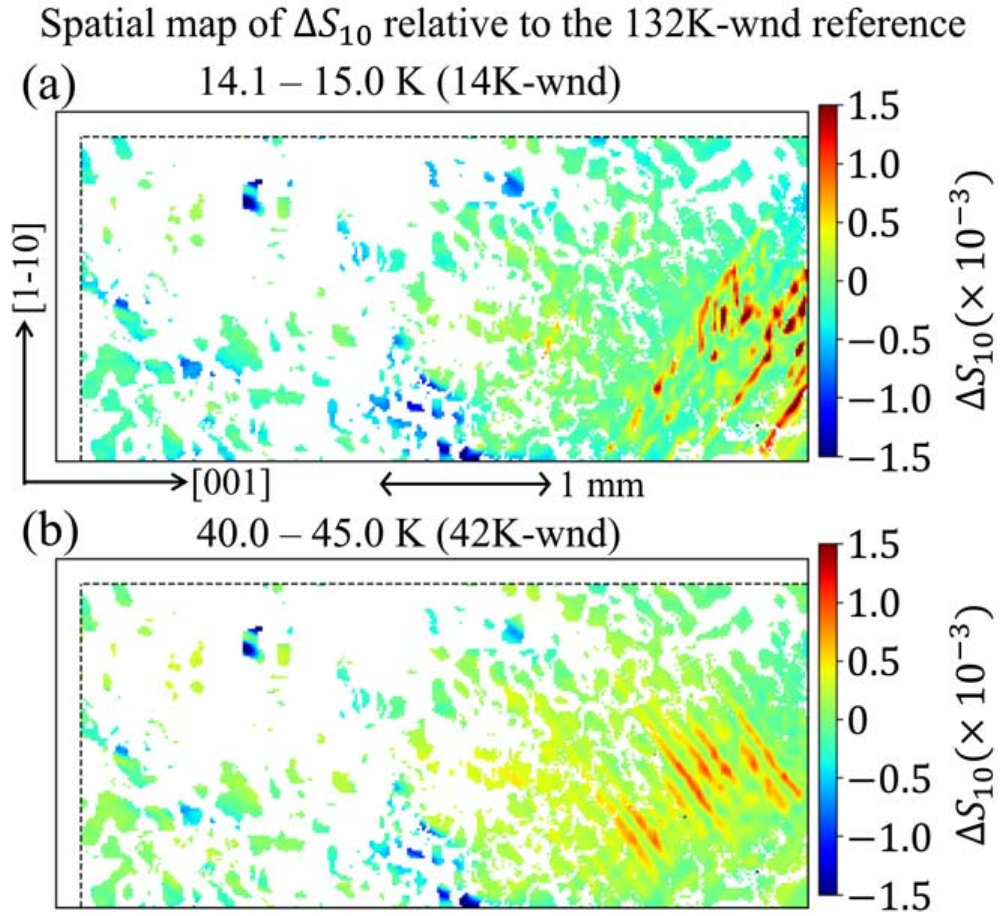}
\end{center}
\caption{(Color online) Maps of $\Delta S_{10}^{(5W)}(x,y)$, defined in Eq.~(9), for (a) the 14K-wnd and (b) the 42K-wnd. As in Fig.~1(b), regions above and to the left of the dashed lines are not shown, as they are excluded from the analysis.}
\end{figure}

\clearpage

\renewcommand{\thefigure}{S\arabic{figure}} 
\renewcommand{\theequation}{S\arabic{equation}} 
\renewcommand{\thetable}{S\arabic{table}} 

\setcounter{equation}{0}
\setcounter{figure}{0}

\maketitle 

\section*{Supplementary Materials} 

\vspace{0.5cm}

\section*{S1.~Detailed construction of the holonomy map} 

\vspace{0.5cm}

The holonomy analysis uses the same birefringence-imaging dataset employed to infer the spatial map of $T_{\rm F}$ in Fig.~1(a). In the experiment, a uniaxial stress of 231~MPa was applied along [001] of SrTiO$_3$ during cooling, and the transmitted polarization state was recorded as a Stokes vector ($\boldsymbol{S} = (S_1, S_2, S_3)$). Measurements were performed at three incident wavelengths, $\lambda=523$, 543, and 575~nm. Because all wavelengths yield qualitatively similar results and the 575-nm dataset provides the highest signal-to-noise ratio, only the 575-nm data are used in the present holonomy analysis.

In continuum elasticity, incompatible strain fields can generate residual rotations upon transport along closed paths, as in the Saint--Venant compatibility framework. Because birefringence imaging reflects strain-induced optical anisotropy, such incompatibility may manifest as path-dependent (non-integrable) structures in the reconstructed optical director field. In this framework, the optical director represents the local orientation of birefringence and thus the direction of strain-induced optical anisotropy in real space. The holonomy measure used here provides a geometric diagnostic of whether the locally reconstructed optical director field remains consistent along closed paths. It may respond to strain-related incompatibility or strain-gradient-induced electromechanical effects that influence the spatial organization of the birefringence field. This viewpoint is conceptually related to the closure-defect picture used in dislocation theory, although no direct correspondence is established here. A rigorous connection between the holonomy of the optical director field and the elastic incompatibility tensor is beyond the scope of this work and remains an important direction for future study.

With this physical background, we describe the detailed construction of the holonomy map. The image size is $302~\times~140$ pixels. As in the main text, the lower-right corner is taken as the origin of the real-space coordinate system, with the positive $x$ direction pointing leftward and the positive $y$ direction upward. Thus, the pixel coordinate $(x,y)$ is defined with respect to this origin. This convention is adopted for two reasons: (i) regions with relatively high $T_{\rm F}$ are concentrated in the lower-right part of the image, and (ii) the square loops used for holonomy calculation are defined from a lower-right base point. For each temperature ($T$) and pixel $(x,y)$, the outgoing polarization state is represented by the Stokes vector:

\begin{equation} 
\boldsymbol{S}^{(T)}(x,y) := \bigl( S^{(T)}_1(x,y), S^{(T)}_2(x,y), S^{(T)}_3(x,y)\bigr)^{\mathsf T}, 
\end{equation}

\noindent where the superscript $(T)$ denotes evaluation at $T$. This notation is used throughout the Supplementary Materials for frame-wise quantities prior to $T$-window aggregation. The instrument output of the Stokes vector is already corrected and normalized under the assumption of a fully polarized state, such that $\|\boldsymbol{S}^{(T)}(x,y)\|\simeq 1$ holds to good accuracy. We nevertheless define the corresponding unit director:

\begin{equation} 
\boldsymbol{n}^{(T)}(x,y) = \frac{\boldsymbol{S}^{(T)}(x,y)}{\|\boldsymbol{S}^{(T)}(x,y)\|}, \label{eq:S1_director_def} 
\end{equation}

\noindent so that only the orientation of the polarization state on the Poincar\'e sphere is retained. Because birefringence imaging does not distinguish $\boldsymbol{n}^{(T)}(x,y)$ and $-\boldsymbol{n}^{(T)}(x,y)$ once the optical retardance exceeds $\lambda/2$, the orientational field is treated as a line field (director field) rather than a vector field. Accordingly, the director is defined on the real projective plane $\mathbb{R}P^2$, with the identification

\begin{equation} 
\boldsymbol{n}^{(T)}(x,y)\sim -\boldsymbol{n}^{(T)}(x,y). 
\end{equation}

\noindent Importantly, this passage from $S^2$ to $\mathbb{R}P^2$ does not by itself imply topological non-triviality. To confirm that the analyzed line field admits a globally consistent lift to $S^2$ (i.e., a consistent global sign choice), we perform a numerical sign-consistency check over the full dataset.

For each $T$ and loop size $L=2$--20, we count pixels within the valid holonomy mask that would require a sign assignment incompatible with a globally consistent lift. No such pixels are found for any $T$ or tested loop size. Thus, no obstruction to a globally consistent sign choice is detected within the sampled region, indicating that the analyzed $\mathbb{R}P^2$ field is topologically trivial on this domain. Consequently, nearest-neighbor minimal rotations can be defined without introducing branch cuts or singular pixels. Because the director is defined only up to sign, the shortest rotation between neighboring pixels must respect the $\mathbb{R}P^2$ identification $\boldsymbol{n}\sim-\boldsymbol{n}$. Accordingly, the local sign of the neighboring director is chosen to minimize the angular separation between the two representatives. 

For a nearest-neighbor step in the $+x$ direction, $(x,y)\to(x+1,y)$, we define 

\begin{equation} 
s_x^{(T)}(x,y) = \mathrm{sign} \left[ \boldsymbol{n}^{(T)}(x,y)\cdot \boldsymbol{n}^{(T)}(x+1,y) \right] \in\{+1,-1\}, \label{eq:S1_sx} 
\end{equation} 

\noindent and use the endpoint representative 

\begin{eqnarray} 
&&\boldsymbol{n}^{\prime(T)}(x+1,y) = s_x^{(T)}(x,y)\, \boldsymbol{n}^{(T)}(x+1,y), \\
&&\boldsymbol{n}^{(T)}(x,y)\cdot \boldsymbol{n}^{\prime(T)}(x+1,y)\ge 0. 
\end{eqnarray} 

\noindent When the scalar product vanishes exactly, we set $s_x^{(T)}(x,y)=+1$. An analogous construction applies in the $+y$ direction, $(x,y)\to(x,y+1)$, yielding $s_y^{(T)}(x,y)$ and $\boldsymbol{n}^{\prime(T)}(x,y+1)$. The shortest rotation between neighboring directors is represented by a unit quaternion. For the $+x$ direction, we define

\begin{align} 
\boldsymbol{u}^{(T)}_x(x,y) &= \boldsymbol{n}^{(T)}(x,y)\times \boldsymbol{n}^{\prime(T)}(x+1,y), \\ 
w^{(T)}_x(x,y) &= 1+ \boldsymbol{n}^{(T)}(x,y)\cdot \boldsymbol{n}^{\prime(T)}(x+1,y), \\ 
q^{(T)}_x(x,y) &= \frac{ \bigl( w^{(T)}_x(x,y),\, \boldsymbol{u}^{(T)}_x(x,y) \bigr) }{ \sqrt{ \bigl(w^{(T)}_x(x,y)\bigr)^2+ \|\boldsymbol{u}^{(T)}_x(x,y)\|^2 } }. \label{eq:S1_qx} 
\end{align} 

\noindent The same construction yields $q^{(T)}_y(x,y)$ for the $+y$ direction. Reverse directions are represented by inverse quaternions:

\begin{equation} 
q^{(T)}_{-x}(x,y)= \bigl(q^{(T)}_x(x-1,y)\bigr)^{-1}, \qquad q^{(T)}_{-y}(x,y)= \bigl(q^{(T)}_y(x,y-1)\bigr)^{-1}. 
\end{equation} 

\noindent For compact notation, we denote by $q^{(T)}_e$ the unit quaternion associated with an oriented nearest-neighbor edge $e$, so that $q^{(T)}_{-e}=(q^{(T)}_e)^{-1}$ by construction. This edge-based formulation also clarifies that the construction is not tied to the two-dimensional image plane. In the present experiment, the available data are sampled on an $xy$ pixel grid. Therefore, the relevant oriented edges are the nearest-neighbor bonds in the $\pm x$ and $\pm y$ directions. If three-dimensional director data were available, minimal-rotation quaternions could be assigned analogously to oriented nearest-neighbor edges in the $\pm x$, $\pm y$, and $\pm z$ directions. More generally, for any closed path $\gamma$ on such a spatial sampling graph, a loop quaternion can be defined as

\begin{equation}
Q_{\gamma}^{(T)}=\prod_{e\in\overrightarrow{\gamma}} q_e^{(T)},
\end{equation}

\noindent where the product is taken in the traversal order of the path. Because each $q_e^{(T)}$ is a unit quaternion representing a rotation acting on the three-component director, $Q_{\gamma}^{(T)}$ is also a unit quaternion regardless of the dimensionality of the real-space sampling graph. The square-loop definition used below is a particular implementation of this general closed-path construction, adopted for the two-dimensional birefringence images analyzed in this work.

In this two-dimensional implementation, the holonomy angle is evaluated from the product of edge rotations around a finite square loop. For a fixed loop size $L$, let $\Box_L(x,y)$ denote the $L~\times~L$-pixel square with lower-right corner at the base point $(x,y)$. The ordered boundary of this loop is traversed from the lower-right base point $(x,y)$ in the sequence
\[
+L\ \text{steps in } +x \;\rightarrow\; +L\ \text{steps in } +y \;\rightarrow\; +L\ \text{steps in } -x \;\rightarrow\; +L\ \text{steps in } -y,
\]
returning to $(x,y)$. For this square loop, the corresponding loop quaternion is written as
\begin{equation}
Q_L^{(T)}(x,y) = \prod_{e\in\overrightarrow{\partial\Box_L(x,y)}} q_e^{(T)} \in \mathrm{SU}(2). \label{eq:S1_Qk}
\end{equation}

\noindent Because quaternion multiplication is non-commutative, the traversal order is essential. Here the local rotations are represented by unit quaternions, meaning the loop product is naturally written in $\mathrm{SU}(2)$, which is the double cover of $\mathrm{SO}(3)$.

If the orientational field admits a globally consistent lift (i.e., path-integrable), the accumulated rotation around the loop is trivial. In contrast, if a geometric inconsistency is enclosed, a nontrivial residual rotation remains after traversal. Writing the loop quaternion in scalar--vector form,

\begin{equation} 
Q_L^{(T)}(x,y) = \bigl( w_L^{(T)}(x,y),\, \boldsymbol{v}_L^{(T)}(x,y) \bigr), 
\end{equation} 

\noindent where $w_L^{(T)}(x,y)\in\mathbb{R}$ and $\boldsymbol{v}_L^{(T)}(x,y)\in\mathbb{R}^3$ are the scalar and vector parts of the residual loop quaternion, respectively. Equivalently, for a unit quaternion,

\begin{equation} Q_L^{(T)}(x,y) = \left( \cos\frac{\omega_L^{(T)}(x,y)}{2},\, \hat{\boldsymbol{u}}_L^{(T)}(x,y) \sin\frac{\omega_L^{(T)}(x,y)}{2} \right), \label{13} 
\end{equation}

\noindent where $\hat{\boldsymbol{u}}_L^{(T)}(x,y)$ is the unit rotation axis ($\hat{\boldsymbol{u}}_L^{(T)}(x,y)=\boldsymbol{v}_L^{(T)}(x,y)/\|\boldsymbol{v}_L^{(T)}(x,y)\|$). Using the identification $Q_L^{(T)}(x,y)\sim -Q_L^{(T)}(x,y)$, we select the canonical representative satisfying 

\begin{equation} 
w_L^{(T)}(x,y)\ge 0. 
\end{equation} 

\noindent The holonomy angle is then defined as 

\begin{equation} 
\omega_L^{(T)}(x,y) = 2 \arctan2 \left( \|\boldsymbol{v}_L^{(T)}(x,y)\|, \,w_L^{(T)}(x,y) \right) \in[0,\pi], \label{eq:S1_omega_def} 
\end{equation} 

\noindent where $\omega_L^{(T)}(x,y)=0$ corresponds to a vanishing loop residual, whereas larger values indicate stronger path-dependent non-integrability of the locally stitched orientational field. In the main analysis, we set $L=10$, motivated by previously reported real-space scales in the same sample, namely the $\sim$10-pixel width of [111]-oriented slip-plane structures under stress and the comparable scale of ferroelectric textures. Preliminary calculations for $L=2$--20 yield qualitatively similar spatial trends, indicating that the results are robust to the choice of loop size. A systematic study of the $L$ dependence is deferred to future work. When the loop is scanned with a unit stride over the image, the nominal $10\times10$-pixel loop extends beyond the field of view near the left and top boundaries. In these regions, the loop size is reduced to the largest admissible value,

\begin{equation} 
L_{\rm eff}(x,y) = \min\bigl(L,\, L_x^{\max},\, L_y^{\max} \bigr), 
\end{equation} 

\noindent where $L_x^{\max}$ and $L_y^{\max}$ denote the maximum admissible lengths along the $+x$ and $+y$ directions from the base point $(x,y)$, respectively. The holonomy angle is then computed using $L_{\rm eff}(x,y)$ in place of $L$. As a result, the leftmost and uppermost 9-pixel-wide strips are not directly comparable to the interior region, where the nominal loop size is $L=10$, and are therefore used for visualization only. These boundary strips are excluded from the main statistical aggregation. Hereafter, unless otherwise noted, the nominal loop size is fixed at $L=10$. 

The holonomy calculation is implemented in Python~3.11 using standard numerical libraries (e.g., ``numpy''). The computational cost is moderate: an $\omega$ map for the full image ($302 \times 140 =42,280$ pixels) is computed in approximately 10~min on a standard desktop workstation. The implementation relies only on basic linear algebra and quaternion arithmetic, and is therefore readily reproducible by non-specialists. The full dataset comprises 3,362 $T$ points acquired during continuous cooling from 300.0 to 14.1~K.

\begin{figure}[tb] 
\begin{center} 
\includegraphics[width=12cm]{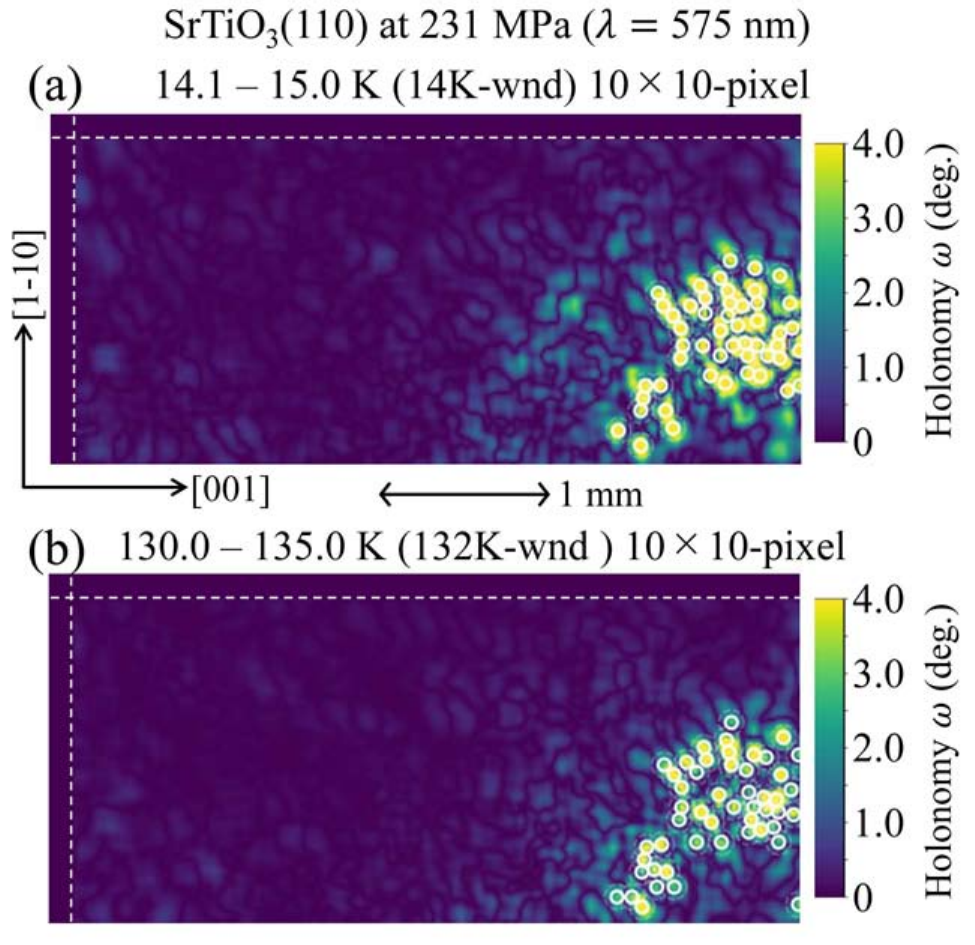} 
\end{center} 
\caption{Representative real-space maps (302$\times$140 pixels) of the holonomy angle $\overline{\omega}_{10}^{(5W)}(x,y)$ computed using $10\times10$-pixel loops, for (a) the 14K-wnd and (b) 132K-wnd. Because the $10\times10$-pixel loop is constructed from the lower-right base point, it extends beyond the field of view near the left and top boundaries. The regions above and to the left of the dashed lines are therefore shown for visualization only. Circles indicate representative local maxima of $\overline{\omega}_{10}^{(5W)}$ extracted by two-dimensional non-maximum suppression (NMS); their size is smaller than the minimum separation scale to clearly mark the peak centers.}
\end{figure} 

\begin{figure}[tb] 
\begin{center} 
\includegraphics[width=12cm]{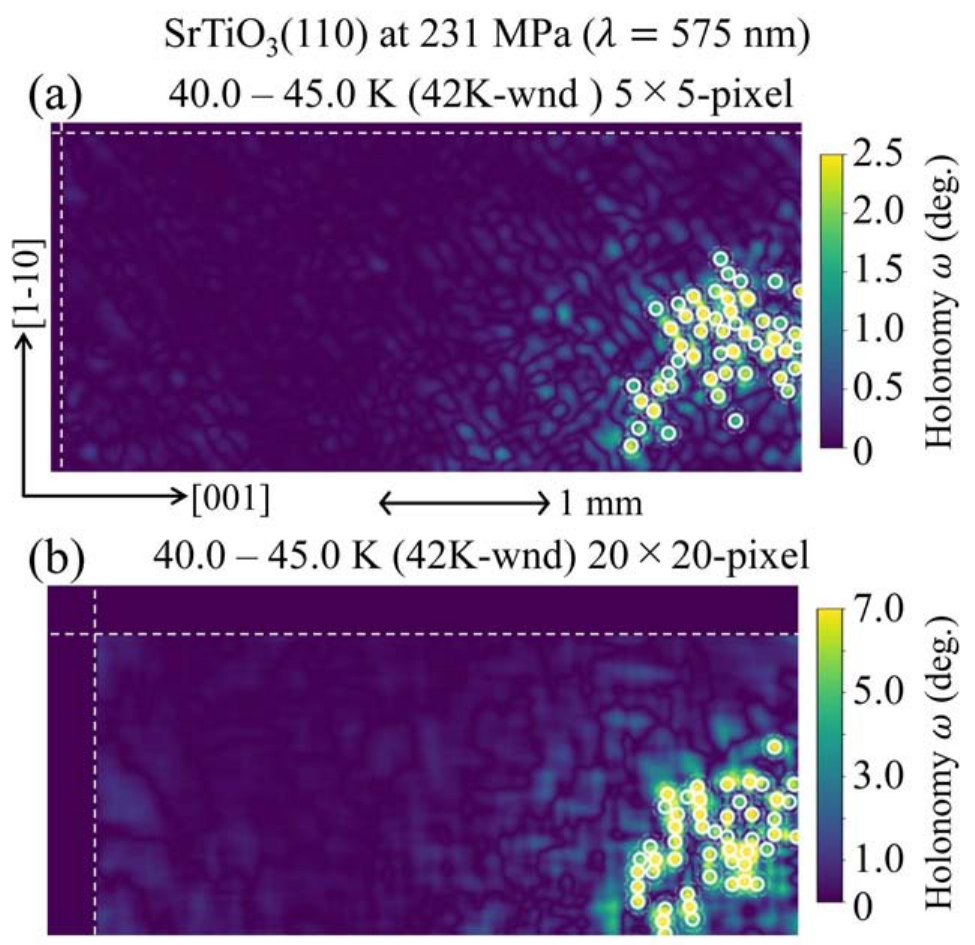} 
\end{center} 
\caption{Representative real-space maps of the holonomy angle $\overline{\omega}_L^{(5W)}(x,y)$ for the 42K-wnd, computed using (a) $5\times5$-pixel loops ($L=5$) and (b) $20\times20$-pixel loops ($L=20$). As in Fig.~S1, regions above and to the left of the dashed lines are shown for visualization only. Circles indicate representative peak positions of $\overline{\omega}_L^{(5W)}$ extracted by NMS.} 
\end{figure} 

For visualization of spatial patterns, the data are grouped into 5-K-wide $T$ windows in descending order from 300.0~K. For a given window ($W$), let $\mathcal{T}_{5W}$ denote the set of $T$ points within that interval. The representative holonomy map for each window is defined as the pixel-wise median:

\begin{equation} 
\overline{\omega}_L^{(5W)}(x,y) := \operatorname{median}_{T\in\mathcal{T}_{5W}} \omega_L^{(T)}(x,y). \label{eq:S1_rep_omega} 
\end{equation}

\noindent Importantly, the holonomy angle is computed independently at each $T$ and only subsequently aggregated over $T$. Thus, the representative map is not obtained by averaging the director field prior to the holonomy calculation. Most 5-K-wide windows contain approximately 60 $T$ points. The lowest-$T$ window, denoted 14K-wnd, spans the interval $[14.1~{\rm K}, 15.0~{\rm K}]$ because the measurement terminates at 14.1~K, and therefore contains only 16 $T$ points.

Figure~S1 shows the representative holonomy maps $\overline{\omega}_{10}^{(5W)}(x,y)$ for the 14K-wnd and the 132K-wnd, $(130.0~{\rm K}, 135.0~{\rm K}]$, which are referred to in the main text (Fig. 1(b)) as representative low- and high-$T$ cases. Because the 14K-wnd contains fewer $T$ points than most other 5-K-wide windows, it is necessary to verify that this reduction does not qualitatively affect the representative map. To this end, the representative-map construction is repeated for higher-$T$ windows using randomly downsampled subsets of $T$ points comparable in size to the 14K-wnd. The resulting maps remain qualitatively stable, and the spatial trends of the extracted peaks and overlap measures vary only weakly. These tests support the conclusion that the representative map of the 14K-wnd shown in Fig.~S1(a) is not significantly biased by the reduced number of available $T$ points.

To identify spatially separated peaks in a representative holonomy map, two-dimensional non-maximum suppression (NMS) is applied. Starting from the global maximum, candidate peaks within a 5-pixel radius are iteratively suppressed, and up to 50 peaks are extracted. The circles in Figs.~1(b) and S1 indicate the detected peak positions; their size is chosen for visualization only and does not correspond to the suppression radius. The extraction of 50 peaks provides a representative set of spatially separated maxima across the field of view. The qualitative spatial trends remain largely unchanged when this number is varied within a reasonable range.

Figure~S2 shows the holonomy maps $\overline{\omega}_{L}^{(5W)}$ for the 42K-wnd, $(40.0~{\rm K}, 45.0~{\rm K}]$, calculated using loop sizes of $5\times5$ pixels ($L=5$) and $20\times20$ pixels ($L=20$). In general, the holonomy angle increases with loop area, reflecting the cumulative rotation along the loop. Consistent with this expectation, the overall spatial patterns of $\overline{\omega}_{L}^{(5W)}$ are qualitatively similar for both loop sizes. For $L=5$, the spatial features appear relatively sharp, whereas for $L=20$, the structures are somewhat smoother due to the larger loop area. Nevertheless, the global spatial tendencies remain largely unchanged, indicating that the qualitative features of the holonomy map are robust to the choice of loop size. Because the loop is defined from the lower-right base point, increasing $L$ slightly shifts the apparent peak positions toward the base-point direction. This effect affects only the precise peak locations and does not alter the global spatial pattern of the holonomy map.

\vspace{1cm}

\section*{S2.~Gradient-based comparison and temperature-bin statistics} 

\vspace{0.5cm}

To compare with the holonomy map, we construct a local-gradient metric, denoted ``grad,'' which quantifies the nearest-neighbor angular variation of the director field on $\mathbb{R}P^2$. For two neighboring pixels $(x,y)$ and $(x',y')$, the sign-invariant angular difference is defined as 

\begin{equation} 
\Delta^{(T)} \bigl((x,y),(x',y')\bigr) := \arccos \left( \left| \boldsymbol{n}^{(T)}(x,y)\cdot \boldsymbol{n}^{(T)}(x',y') \right| \right) \in[0,\pi/2]. \label{eq:S2_delta} 
\end{equation}

\noindent Here, the absolute value accounts for the identification $\boldsymbol{n}\sim-\boldsymbol{n}$, ensuring that the angular difference is evaluated on the real projective plane $\mathbb{R}P^2$. Using the set $\mathcal{N}_{\rm NN}(x,y)$ of valid nearest-neighbor pixels of $(x,y)$, we define 

\begin{figure}[tb] 
\begin{center} 
\includegraphics[width=12cm]{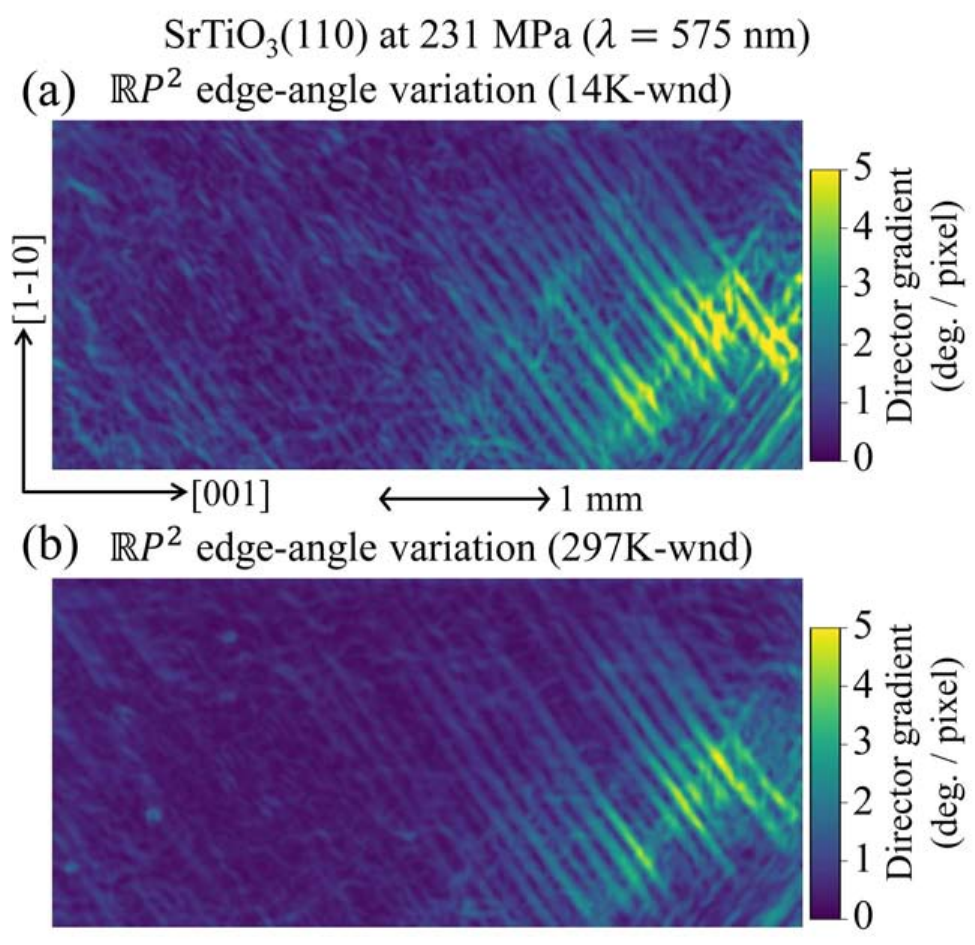} 
\end{center} 
\caption{Representative real-space maps of the nearest-neighbor $\mathbb{R}P^2$ edge-angle variation, $\overline{{\rm grad}}^{(5W)}(x,y)$, for (a) the 14K-wnd and (b) the 297K-wnd, $(295.0~\mathrm{K},300.0~\mathrm{K}]$.} 
\end{figure}

\begin{figure}[tb] 
\begin{center} 
\includegraphics[width=12cm]{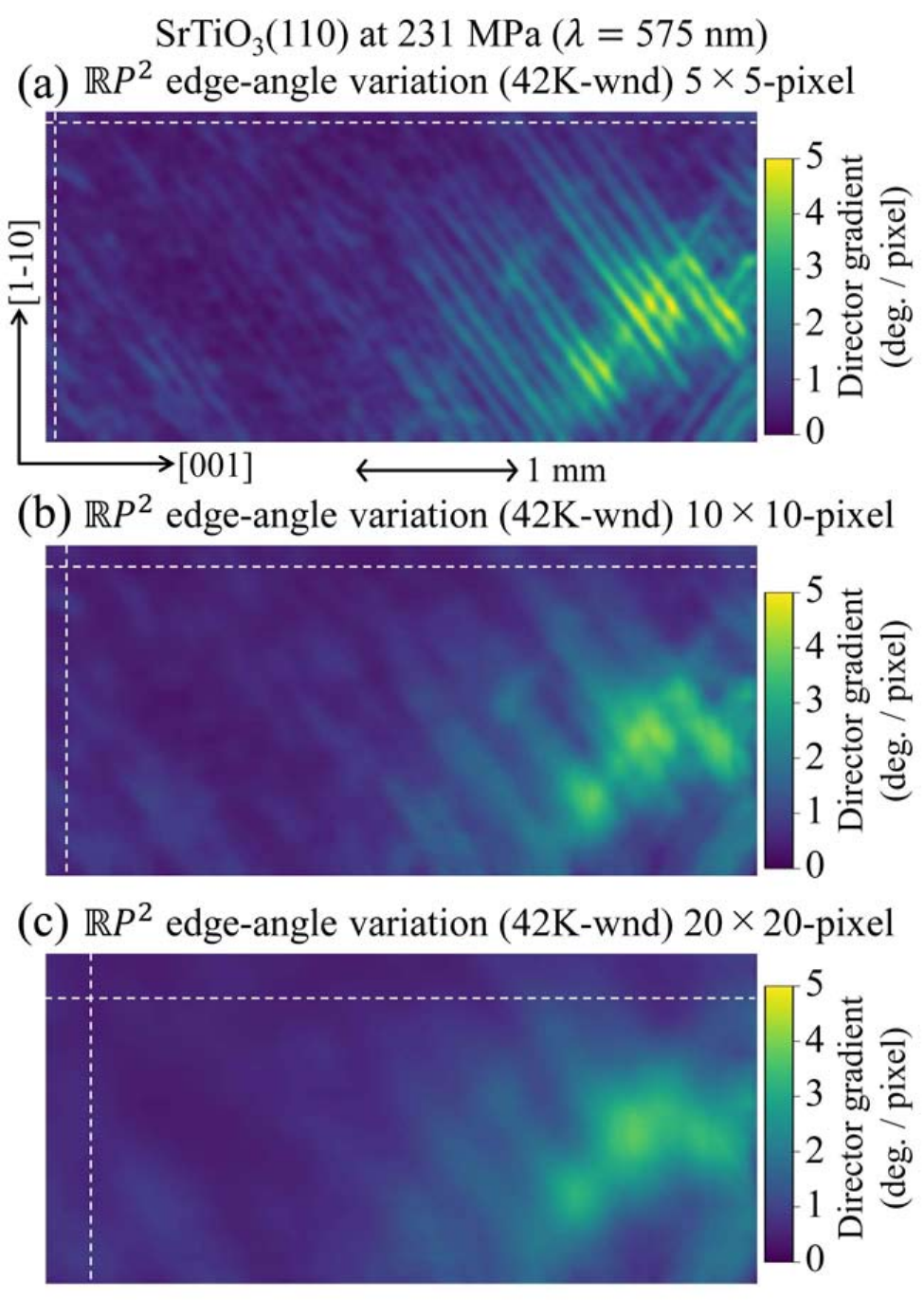} 
\end{center} 
\caption{Representative real-space smoothed grad maps of the nearest-neighbor $\mathbb{R}P^2$ edge-angle variation, $\overline{{\rm grad}}_R^{(5W)}(x,y)$, for the 42K-wnd, spatially averaged over square regions of size (a) $5\times5$, (b) $10\times10$, and (c) $20\times20$ pixels. As in Fig.~S1, regions above and to the left of the dashed lines are shown for visualization only.}
\end{figure}

\begin{equation} 
{\rm grad}^{(T)}(x,y) := \frac{1}{|\mathcal{N}_{\rm NN}(x,y)|} \sum_{(x',y')\in\mathcal{N}_{\rm NN}(x,y)} \Delta^{(T)} \bigl((x,y),(x',y')\bigr). \label{eq:S2_grad} 
\end{equation} 

\noindent Thus, ${\rm grad}^{(T)}(x,y)$ is computed as the mean of the valid incident edge angles. Interior pixels use up to four incident edges, whereas boundary pixels use only the edges available within the image domain. In the present dataset, every analyzed pixel retains at least two valid neighboring edges, so no pixel is undefined in the grad map. The representative grad map for a 5-K-wide window is obtained via the same pixel-wise median aggregation used for the representative holonomy map:

\begin{equation}
\overline{{\rm grad}}^{(5W)}(x,y) :=
\operatorname{median}_{T\in\mathcal{T}_{5W}} {\rm grad}^{(T)}(x,y).
\label{eq:S2_grad_rep}
\end{equation}

\noindent As discussed in the main text for the 42K-wnd, both $\overline{\omega}_{10}^{(5W)}(x,y)$ and $\overline{{\rm grad}}^{(5W)}(x,y)$ are enhanced in regions of high inferred $T_{\rm F}$, but the two maps are not identical. Because holonomy is evaluated over a $10\times10$-pixel loop, its effective spatial resolution is lower than that of the nearest-neighbor grad map, so the boundaries in the $\omega$ map appear more blurred. In particular, the elongated stripe-like feature extending toward the top edge is more pronounced in the grad map than in $\omega$, supporting the view that holonomy is not simply a spatially smoothed surrogate of local gradients but retains additional information related to path dependence and local non-integrability. For comparison across a broader $T$ range, Fig.~S3 shows the representative grad maps for the 14K-wnd and 297K-wnd, $(295.0~\mathrm{K},300.0~\mathrm{K}]$, whose overall spatial tendency remains qualitatively similar across the analyzed $T$ range despite changes in signal magnitude.

To facilitate comparison with the $10\times10$-pixel holonomy map shown in Fig.~1(b), Fig.~S4 presents smoothed grad maps ($\overline{{\rm grad}}_R^{(5W)}(x,y)$) for the same $T$ window, obtained by spatial averaging over square regions of size $R\times R$ pixels with $R=5, 10$, and 20. In each case, the averaging window is shifted with a unit stride to construct the smoothed grad maps. As the averaging area increases, the boundaries become more blurred, reflecting the reduced spatial resolution. Nevertheless, the elongated high-gradient region extending toward the upper edge of the image remains visible in all cases. These results confirm that the characteristic stripe-like structure observed in the grad map is robust with respect to the spatial smoothing scale.

To quantify similarities and differences among the maps in Fig.~1, we compare their high-value regions using the IoU and Dice coefficients. For a map $M(x,y)$, let $\Omega_{p\%}(M)$ denote the set of pixels corresponding to the upper $p\%$ of values of $M$. Then, for two maps $M_1$ and $M_2$, the overlap measures are defined as

\begin{table}[tb] 
\centering 
\caption{Intersection-over-union (IoU) and Dice coefficients between the grad and holonomy maps for different smoothing scales of grad ($R$) and loop sizes of holonomy ($L$) for the 42K-wnd. The overlap is computed between the sets of pixels corresponding to the upper $p\%$ of each map ($p=50\%, 10\%$).}
\begin{tabular}{cccccc||cccccc} 
\hline 
& & \multicolumn{2}{c}{${\rm IoU}_{p\%}$} & \multicolumn{2}{c||}{${\rm Dice}_{p\%}$} & & & \multicolumn{2}{c}{${\rm IoU}_{p\%}$} & \multicolumn{2}{c}{${\rm Dice}_{p\%}$} \\
$R$ & $L$ & 50\% & 10\% & 50\% & 10\% & $R$ & $L$ & 50\% & 10\% & 50\% & 10\% \\ \hline
1 & 5 & 0.562 & 0.431 & 0.719 & 0.602 & 10 & 5 & 0.582 & 0.415 & 0.736 & 0.586 \\
1 & 10 & 0.534 & 0.391 & 0.697 & 0.562 & 10 & 10 & 0.576 & 0.406 & 0.731 & 0.578 \\ 
1 & 20 & 0.506 & 0.363 & 0.672 & 0.533 & 10 & 20 & 0.553 & 0.375 & 0.712 & 0.545 \\ 
5 & 5 & 0.582 & 0.443 & 0.736 & 0.614 & 20 & 5 & 0.594 & 0.376 & 0.745 & 0.546 \\ 
5 & 10 & 0.565 & 0.424 & 0.722 & 0.596 & 20 & 10 & 0.580 & 0.365 & 0.734 & 0.535 \\ 
5 & 20 & 0.542 & 0.389 & 0.703 & 0.560 & 20 & 20 & 0.547 & 0.324 & 0.707 & 0.490 \\ 
\hline 
\end{tabular} 
\end{table}

\begin{eqnarray} 
{\rm IoU}_{p\%}(M_1,M_2)
&=&\frac{|\Omega_{p\%}(M_1)\cap \Omega_{p\%}(M_2)|}
{|\Omega_{p\%}(M_1)\cup \Omega_{p\%}(M_2)|}, \\
{\rm Dice}_{p\%}(M_1,M_2)
&=&
\frac{2|\Omega_{p\%}(M_1)\cap \Omega_{p\%}(M_2)|}
{|\Omega_{p\%}(M_1)|+|\Omega_{p\%}(M_2)|}. 
\end{eqnarray}

\noindent The overlap between the $T_{\rm F}$, holonomy, and grad maps is summarized in the main text. To further examine the robustness of this comparison, we evaluate the IoU and Dice coefficients under different calculation conditions for the 42K-wnd (Table~S1). The analysis is performed by varying the spatial averaging scale of the grad map ($R$), and the loop size used in the holonomy calculation ($L$). For the upper 50\% regions, the IoU values range from 0.51 to 0.59 and the Dice coefficients from 0.67 to 0.75, indicating moderate spatial agreement. Restricting the analysis to the upper 10\% regions reduces the overlap, with IoU values of 0.32--0.44 and Dice coefficients of 0.49--0.61. Varying the smoothing scale $R$ does not substantially change these metrics, indicating that the results are qualitatively robust. These results demonstrate that the spatial pattern of the holonomy map cannot be explained solely by simple smoothing of the gradient field, but instead reflects additional information arising from path-dependent rotation accumulation and the non-integrability of the orientational field.

\begin{table}[tb] 
\centering 
\caption{Pearson and Spearman correlation coefficients between $\overline{{\rm grad}}^{(5W)}(x,y)$ and $\overline{\omega}_{10}^{(5W)}(x,y)$ for the 42K-wnd, corresponding to $R=1$ and $L=10$, respectively. Here $\alpha_{\rm grad}$ and $\alpha_{\omega}$ denote selection criteria corresponding to the upper $p\%$ of pixels in the grad and holonomy maps, respectively. All selections are defined only within the valid support of the $L=10$ holonomy map, and the correlations are evaluated over the intersection of the corresponding selected pixel sets.}
\begin{tabular}{ccccccr}
\hline
$R$ & $L$ & $\alpha_{\rm grad}$ & $\alpha_{\omega}$ & Pearson & Spearman& No. of pixels \\
\hline
1 & 10 & all & all & 0.632 & 0.613  & 38,383 \\
1 & 10 & all & 50\% & 0.580 & 0.631  & 19,192 \\
1 & 10 & all & 10\% & 0.334 & 0.332  & 3,838 \\
1 & 10 & 50\% & all & 0.564 & 0.583  & 19,192 \\
1 & 10 & 10\% & all & 0.342 & 0.349  & 3,838 \\
1 & 10 & 50\% & 50\% & 0.528 & 0.546  & 13,997 \\
1 & 10 & 10\% & 10\% & 0.255 & 0.287 & 2,114 \\
\hline
\end{tabular} 
\end{table}

To further quantify the relationship between the grad and holonomy maps, we computed the correlation coefficients between their pixel values (Table~S2). Here, $\alpha_{\rm grad}$ and $\alpha_{\omega}$ denote selection criteria corresponding to fixed upper-percentile regions of the grad and holonomy maps, respectively. The correlations are computed over pixels that satisfy both $\alpha_{\rm grad}$ and $\alpha_{\omega}$. When all valid pixels are included, the Pearson correlation coefficient is $r\simeq 0.63$, and the Spearman (rank) correlation coefficient is $r\simeq 0.61$, indicating a moderate positive correlation. Restricting the analysis to regions with large holonomy values reduces the correlation but it remains finite; within the upper 10\% of pixels in the holonomy map, the Pearson coefficient is $r\simeq 0.33$, suggesting that regions of high holonomy are often associated with relatively large gradient values. Similarly, restricting the analysis to the upper 10\% of pixels in the grad map gives a Pearson coefficient of $r\simeq 0.34$. When both selections are simultaneously restricted to their upper 10\% regions, the Pearson coefficient decreases further to $r\simeq 0.26$. These results indicate a partial spatial overlap between the two quantities. At the same time, they suggest that holonomy is not simply a smoothed representation of the gradient field or a measure of local gradient magnitude, but instead captures, at least in part, the accumulation of path-dependent rotations in the orientational field, i.e., local non-integrability.

The statistical uncertainty of the correlation coefficient is estimated using Fisher's $z$ transformation, which provides a good approximation for both Pearson and Spearman correlations for large sample sizes. For the upper 10\% region in both $\alpha_{\rm grad}$ and $\alpha_{\omega}$, the number of selected pixels is 2,114, corresponding to a nominal uncertainty of approximately $\pm 0.04$ in $r$ at the 95\% confidence level under the assumption of independent samples. In practice, neighboring pixels are spatially correlated, so the effective number of independent samples is smaller and the true uncertainty is likely larger than this estimate. Nevertheless, the observed differences in $r$ are substantially larger than this scale and do not affect the qualitative conclusions.

\begin{figure}[tb] 
\begin{center} 
\includegraphics[width=12cm]{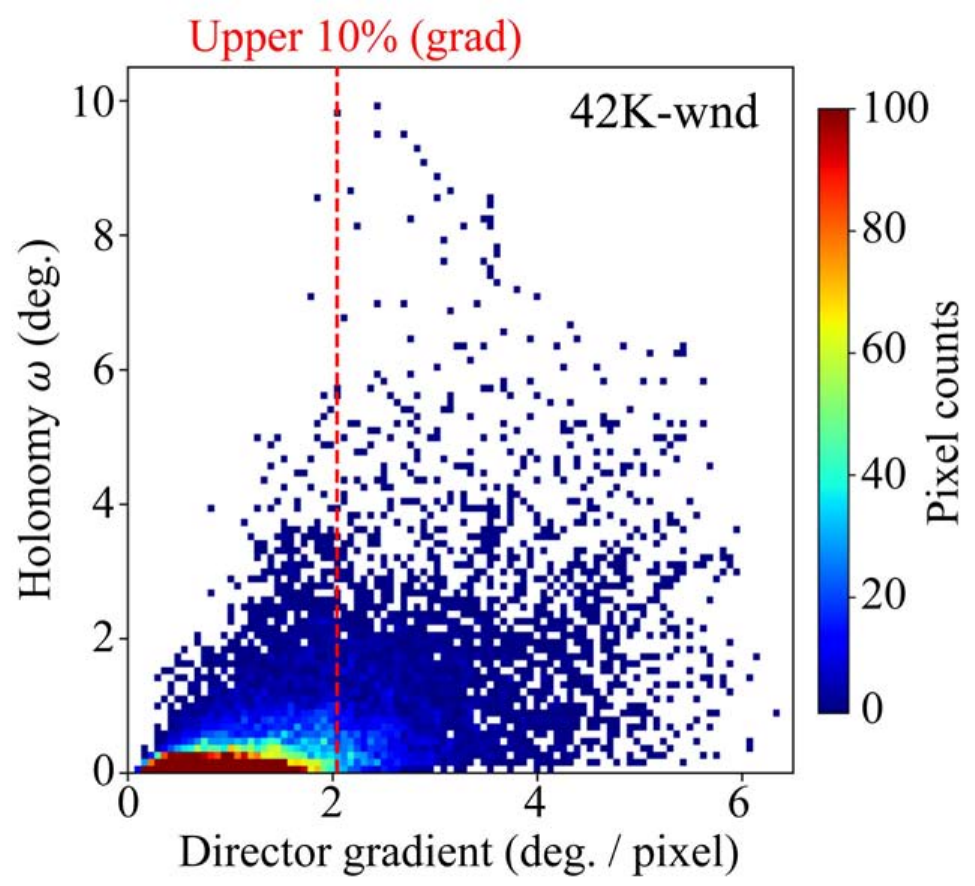} 
\end{center} 
\caption{Two-dimensional histogram showing the pixel-wise relationship between $\overline{{\rm grad}}^{(5W)}(x,y)$ and $\overline{\omega}_{10}^{(5W)}(x,y)$ for the 42K-wnd. Each bin counts the number of pixels within the corresponding grad and holonomy ranges. A positive correlation between the two quantities is evident, consistent with the correlation coefficients reported in Table~S2. The red dashed vertical line indicates the threshold corresponding to the upper 10\% of pixels in the grad map.}
\end{figure}

To visualize the relationship between grad ($R=1$) and holonomy ($L=10$) for the 42K-wnd, we constructed a two-dimensional histogram of their pixel values (Fig.~S5). The horizontal axis represents $\overline{{\rm grad}}^{(5W)}$, the vertical axis represents $\overline{\omega}_{10}^{(5W)}$, and the color encodes the number of pixels in each bin. The red dashed line indicates the threshold corresponding to the upper 10\% of pixels in grad. In regions with small grad, holonomy values are strongly concentrated near zero. As grad increases, the upper bound of holonomy gradually expands, producing a triangular (fan-shaped) distribution. This pattern reflects the definition of holonomy as a loop integral of local orientational changes: while its upper bound is influenced by the magnitude of the local gradient, the actual value depends on the spatial arrangement of the orientational field.

A particularly important observation is that even in the region to the right of the red dashed line, corresponding to the upper 10\% of pixels in grad, holonomy values remain broadly distributed. Many pixels with large local gradients still exhibit small holonomy, indicating that a large gradient alone is not sufficient to produce strong holonomy. Large holonomy requires both a substantial local gradient and a spatial configuration of the orientational field that generates non-integrable rotational structures. Conversely, regions with large holonomy tend to occur where grad is relatively high, indicating that holonomy is not entirely independent of the local gradient. Nevertheless, because holonomy remains widely distributed even within large-grad regions, the relationship between the two quantities is clearly not a simple proportional one. These results demonstrate that holonomy is not merely a measure of local-gradient magnitude, but rather reflects the path-dependent accumulation of rotations in the orientational field, i.e., the local non-integrability of the field. 

\vspace{1cm}
\clearpage

\section*{S3.~Temperature-window statistics of holonomy}

\vspace{0.5cm}

To examine the $T$ dependence of holonomy at scale $L$, we rebin the full cooling sequence into 1-K-wide $T$ bins and evaluate frame-wise summary statistics. For each $T$, let $\Omega_{L, {\rm all}}^{(T)}$ denote the set of all valid pixels. The spatial median over all valid pixels is defined as

\begin{equation} 
\omega_{L, {\rm all}}^{(T)} := \operatorname{median}_{(x,y)\in\Omega_{L, {\rm all}}^{(T)}} \omega_{L}^{(T)}(x,y). \label{eq:S2_omega_all} 
\end{equation} 

\noindent We further define the subset corresponding to the upper $p\%$ of pixels:

\begin{equation}
\Omega_{L, p\%}^{(T)} :=
\left\{
(x,y)\in\Omega_{L, {\rm all}}^{(T)}
\,\big|\,
\omega_{L}^{(T)}(x,y)
\text{ belongs to the upper } p\% \text{ of pixels}
\right\},
\end{equation}

\noindent with $p=10$ in the present analysis. The corresponding frame-wise median is 

\begin{equation} 
\omega_{L, p\%}^{(T)} := \operatorname{median}_{(x,y)\in\Omega_{L, p\%}^{(T)}} {\omega}_{L}^{(T)}(x,y). \label{eq:S2_omega_top} 
\end{equation} 

\noindent Let $\mathcal{T}_{1W}$ denote the set of $T$ points within a 1-K-wide window. The bin-representative values are defined as the average of the frame-wise medians: 

\begin{equation} 
\overline{\omega}_{L, \alpha}^{(1W)} := \frac{1}{|\mathcal{T}_{1W}|} \sum_{T\in\mathcal{T}_{1W}} \omega_{L, \alpha}^{(T)} \qquad \alpha\in\{{\rm all},10\%\}. \label{eq:S2_omega_bin} 
\end{equation}

\noindent Figure~S6 shows the $T$ dependence of the average holonomy angle evaluated at 50 spatially separated local maxima extracted by NMS, denoted $\overline{\omega}_{\rm 10, NMS}^{(1W)}$. Since this quantity is computed from only 50 peaks, its fluctuations are somewhat larger than those of $\overline{\omega}_{10, 10\%}^{(1W)}$ (Fig. 2(a)). However, its overall $T$ dependence closely follows that of $\overline{\omega}_{10, 10\%}^{(1W)}$. Because the NMS procedure selects pixels with particularly large holonomy, $\overline{\omega}_{\rm 10, NMS}^{(1W)}$ takes systematically higher values than $\overline{\omega}_{10, 10\%}^{(1W)}$. These results indicate that the $T$ dependence of the holonomy signal is robust with respect to the choice of threshold or selection procedure.

\begin{figure}[tb] 
\begin{center} 
\includegraphics[width=12cm]{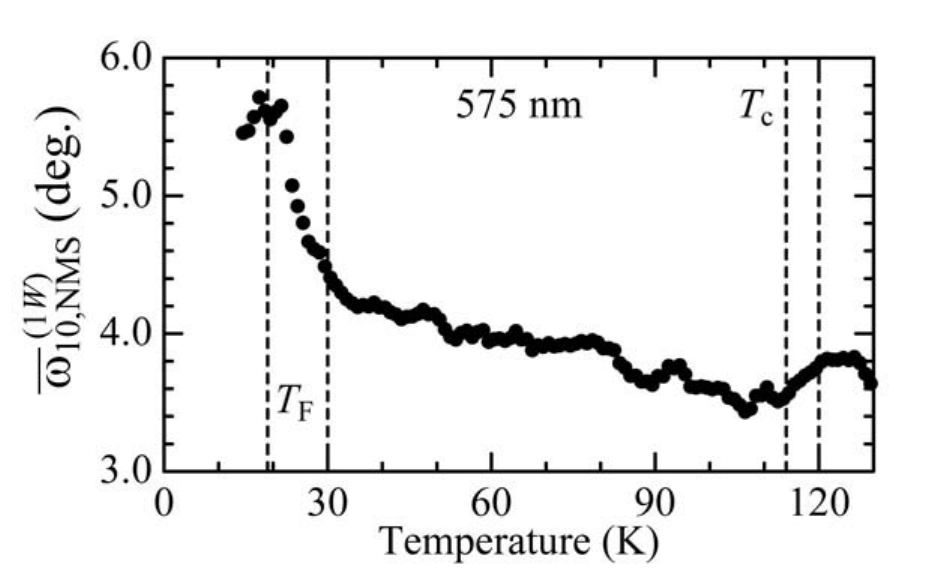} 
\end{center} 
\caption{Temperature dependence of the holonomy angle $\overline{\omega}_{\rm 10, NMS}^{(1W)}$, defined as the average holonomy evaluated at 50 spatially separated local maxima extracted via NMS. Dashed vertical lines indicate the reported ranges of $T_{\rm c}$ and $T_{\rm F}$.} 
\end{figure}

The holonomy map provides not only the residual angle but also the associated rotation axis. Let $\hat{\boldsymbol{u}}_L^{(T)}(x,y)\in\mathbb{R}^3$ denote the axis vector extracted from the loop quaternion at a given $T$, as introduced via the axis--angle representation in Eq.~(\ref{13}). Because $\hat{\boldsymbol{u}}_L^{(T)}(x,y)\sim-\hat{\boldsymbol{u}}_L^{(T)}(x,y)$, a sign-invariant description is required. For a pixel set $\Omega_{L, \alpha}^{(T)}$, we define the unit axis $\hat{\boldsymbol{u}}_L^{(T)}(x,y)$ and the corresponding second-rank alignment tensor:

\begin{equation} 
\boldsymbol{A}_{L, \alpha}^{(T)} := \frac{1}{|\Omega_{L, \alpha}^{(T)}|} \sum_{(x,y)\in\Omega_{L, \alpha}^{(T)}} \hat{\boldsymbol{u}}_L^{(T)}(x,y)~\hat{\boldsymbol{u}}_L^{(T)}(x,y)^{\mathsf T} \in\mathbb{R}^{3\times3}, \label{eq:S2_align_tensor} 
\end{equation} 

\noindent where $\alpha\in\{{\rm all}, 10\%\}$ denotes either all pixels or the subset corresponding to the upper 10\% of $\omega_L^{(T)}(x,y)$. This tensor is symmetric and satisfies 

\begin{equation} 
\mathrm{tr}\,\boldsymbol{A}_{L, \alpha}^{(T)}=1.  \label{eq:S2_align_tensor2}
\end{equation} 

\noindent Let $\lambda_{L,\alpha}^{\max}(T)$ denote the largest eigenvalue of $\boldsymbol{A}_{L, \alpha}^{(T)}$. From Eqs.~(\ref{eq:S2_align_tensor}) and (\ref{eq:S2_align_tensor2}), an isotropic axis distribution yields $\lambda_{L, \alpha}^{\max}(T)=1/3$, whereas perfect alignment along a single direction gives $\lambda_{L,\alpha}^{\max}(T)=1$. We therefore define the frame-wise axis-alignment order parameter by linearly rescaling $\lambda_{L, \alpha}^{\max}(T)$ so that these limiting cases correspond to 0 and 1, respectively: 

\begin{equation} 
S_{L, \alpha}^{(T)} := \frac{3\lambda_{L,\alpha}^{\max}(T)-1}{2}. \label{eq:S2_S_def} 
\end{equation} 

\noindent Thus, $S_{L, \alpha}^{(T)}\simeq1$ indicates strong alignment of the rotation axes, whereas $S_{L, \alpha}^{(T)}\simeq0$ indicates an approximately isotropic distribution. The corresponding 1-K-bin averages are defined analogously to the holonomy statistics:

\begin{equation} 
\overline{S}_{L, \alpha}^{(1W)} := \frac{1}{|\mathcal{T}_{1W}|} \sum_{T\in\mathcal{T}_{1W}} S_{L, \alpha}^{(T)} \qquad \alpha\in\{{\rm all},10\%\}. \label{eq:S2_S_bin} 
\end{equation} 

\noindent As shown in Fig.~2(b) of the main text, $\overline{S}_{10, \alpha}^{(1W)}$ exhibits a characteristic $T$ dependence associated with changes in electromechanical inhomogeneity and domain formation. Here, we examine the robustness of this behavior with respect to the choice of percentile threshold. Figure~S7 shows the $T$ dependences of $\overline{S}_{10, 1\%}^{(1W)}$ and $\overline{S}_{10, 50\%}^{(1W)}$. Despite the different percentile thresholds of the holonomy-angle distribution, the qualitative $T$ dependence remains largely unchanged, indicating that the observed behavior of $\overline{S}_{L, \alpha}^{(1W)}$ is robust with respect to the choice of percentile threshold.

\begin{figure}[tb] 
\begin{center} 
\includegraphics[width=14cm]{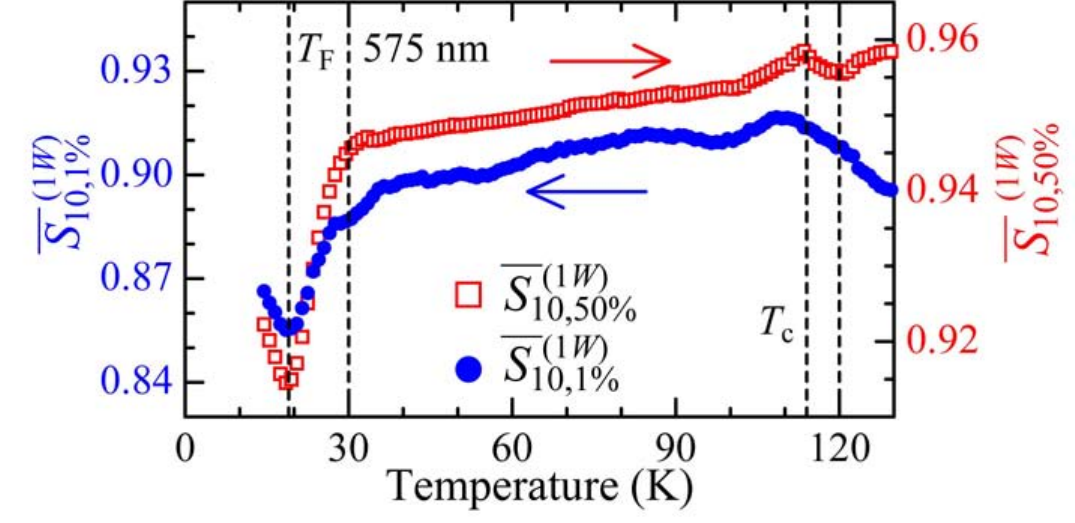} 
\end{center} 
\caption{Temperature dependences of the axis-alignment order parameter $\overline{S}_{10, \alpha}^{(1W)}$. Here, $\alpha={\rm 1\%}$ and $\alpha={\rm 50\%}$ denote averages over the upper 1\% and upper 50\% of pixels in $\omega_{10}^{(T)}(x,y)$, respectively. Dashed vertical lines indicate the reported ranges of $T_{\rm c}$ and $T_{\rm F}$.} 
\end{figure}

The axis-order parameter $S_{L, \alpha}^{(T)}$ used in this study quantifies the spatial alignment of the holonomy rotation axis $\hat{\boldsymbol{u}}_{L}^{(T)}$. Here, we briefly relate this quantity to the distribution of fast-axis orientations obtained from birefringence images. In birefringence imaging, the polarization state of transmitted light at each pixel encodes the local optical anisotropy direction, which can be represented by a director $\boldsymbol{n}^{(T)}(x,y)$ on the Poincar\'e sphere. This director forms a line field satisfying $\boldsymbol{n}\sim -\boldsymbol{n}$ and experimentally reflects the orientation of the fast axis in the birefringence pattern. Figure~S8(a) shows a color map of the fast-axis orientation at 40.0~K. The orientation is nearly uniform across the field of view, indicating strong global alignment of the optical director. Figures~S8(b) and S8(c) show histograms of the fast-axis orientation at 40.0 and 130.0~K, respectively. The distributions are narrow, with standard deviations $\sigma_{\rm 40K}= 2.71^\circ$ and $\sigma_{\rm 130K}= 2.48^\circ$. Strictly speaking, orientation angles should be analyzed using directional statistics (e.g., a von Mises distribution); however, because the angular spread is very small, the conventional linear standard deviation provides an excellent approximation.

The alignment of the optical director can be quantified using directional statistics. Since the director represents a line field on $\mathbb{R}P^2$ and satisfies the nematic symmetry

\begin{equation} 
\theta \sim \theta+\pi, 
\end{equation} 

\noindent the orientation angles are doubled ($2\theta$) so that the $\pi$-periodic director field is treated as a $2\pi$-periodic variable. The mean resultant length of the fast-axis orientation is then calculated as 

\clearpage

\begin{figure}[tb] 
\begin{center} 
\includegraphics[width=12cm]{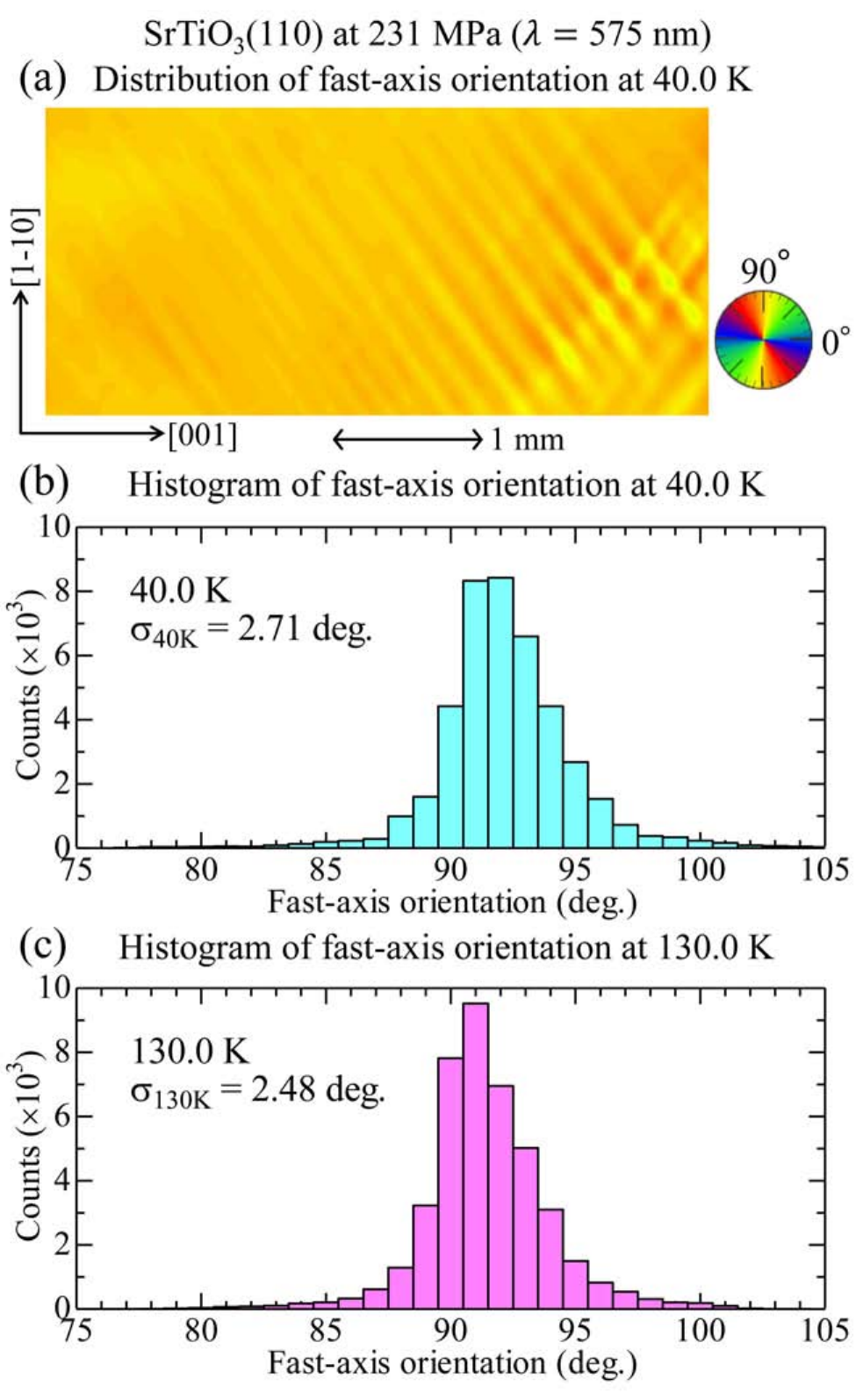}
\end{center}
\caption{(a) Spatial map of the fast-axis orientation extracted from the birefringence image at $40.0$~K. Histograms of the fast-axis orientation at (b) $40.0$~K ($\sigma_{\rm 40K} \simeq 2.71^\circ$) and (c) $130.0$~K ($\sigma_{\rm 130K} \simeq 2.48^\circ$), illustrating the angular dispersion at the two temperatures.}
\end{figure} 

\clearpage

\begin{equation} 
\overline{R}_{\rm fast}=\frac{\sqrt{\left[\sum_{i=1}^{|\mathcal{N}_{\rm fast}|} \cos(2\theta_i)\right]^2+\left[\sum_{i=1}^{|\mathcal{N}_{\rm fast}|} \sin(2\theta_i)\right]^2}}{|\mathcal{N}_{\rm fast}|}, \label{rbar}
\end{equation} 

\noindent where $\mathcal{N}_{\rm fast}$ is the subset corresponding to the fast-axis orientation map and $\theta_i$ is the orientation of the $i$-th pixel among $\mathcal{N}_{\rm fast}$. $\overline{R}_{\rm fast}$ quantifies the degree of alignment of the director field, with $\overline{R}_{\rm fast}\simeq 1$ indicating strong orientational order. While $\overline{R}_{\rm fast}$ provides a directional-statistics measure of alignment, transforming it to the conventional nematic order parameter $S_{\rm fast}$ enables direct comparison with second-rank tensor descriptions of orientational order: 

\begin{equation} 
S_{\rm fast}= \frac{3\overline{R}_{\rm fast}^2-1}{2}, \label{S34}
\end{equation} 

\noindent which yields a scalar measure consistent with standard nematic order parameters. For the fast-axis histogram in Fig.~S8(b), evaluated over all pixels at 40.0~K, this procedure gives

\begin{equation}
\overline{R}_{\rm fast}(40.0~{\rm K})\simeq 0.995, \qquad S_{\rm fast}(40.0~{\rm K})\simeq 0.985.
\end{equation}

\noindent These results confirm that the optical director field is strongly aligned across the sample. Under such conditions, the holonomy rotation axes derived from the director rotations are likewise expected to exhibit global alignment, yielding values of the axis-order parameter $\overline{S}_{10, \alpha}^{(1W)}$ close to unity when all pixels are included. In contrast, in regions with large holonomy angles, the director field becomes geometrically incompatible along closed loops, causing the corresponding rotation axes to be more dispersed. Consequently, when only the high-$\omega$ subset of pixels is considered, the axis-order parameter decreases relative to the global value. In the present dataset, $\overline{S}_{10, \alpha}^{(1W)}$ for $\alpha=1\%$ and 10\%\ is consistently lower than $\overline{S}_{\rm 10, all}^{(1W)}$.

To assess the effect of orientational fluctuations in the director, we rewrite the mean resultant length $\overline{R}_{\rm fast}$ from Eq.~(\ref{rbar}) in a form that explicitly accounts for the nematic symmetry $\theta \sim \theta+\pi$:

\begin{equation}
\overline{R} = \left| \left\langle e^{i2\theta} \right\rangle \right|
\end{equation}

\noindent where $\langle \cdots \rangle$ denotes a spatial average. Assuming weak fluctuations around a mean orientation $\theta_0$, we write

\begin{eqnarray}
\theta &=& \theta_0 + \delta\theta \qquad \langle \delta\theta \rangle = 0, \\
e^{i2\theta} &=& e^{i2\theta_0} e^{i2\delta\theta} \nonumber \\
 &\simeq& e^{i2\theta_0} \left(1 + i 2\delta\theta - 2\delta\theta^2\right),
\end{eqnarray}

\noindent where the exponential is expanded to second order in $\delta\theta$. Averaging and using $\langle \delta\theta \rangle = 0$, we obtain

\begin{eqnarray}
\left\langle e^{i2\delta\theta} \right\rangle &\simeq& 1 - 2\,\langle \delta\theta^2 \rangle, \\
\overline{R}_{\rm fast} &\simeq& \left| e^{i2\theta_0} \left(1 - 2\,\langle \delta\theta^2 \rangle \right) \right| \nonumber \\
 &\simeq& 1 - 2\,\langle \delta\theta^2 \rangle,
\end{eqnarray}

\noindent where the linear term vanishes and $|e^{i2\theta_0}|=1$ is used, assuming $\langle \delta\theta^2 \rangle \ll 1$. Substituting this into Eq.~(\ref{S34}) gives the nematic order parameter:

\begin{equation}
S_{\rm fast} \simeq 1 - 6\,\langle \delta\theta^2 \rangle,
\end{equation}

\noindent showing that deviations of $S_{\rm fast}$ from unity are directly proportional to the mean-square angular fluctuations of the director.

To compare director fluctuations at different temperatures (40.0 and 130.0~K) as shown in Figs.~S8, we decompose the angular fluctuation as

\begin{equation}
\delta\theta_{\rm 40K} = \delta\theta_{\rm 130K} + \delta\theta_{\rm add},
\end{equation}

\noindent where $\delta\theta_{\rm add}$ represents the additional fluctuation relative to the high-$T$ reference (130.0~K) and is assumed statistically independent of $\delta\theta_{\rm 130K}$. Under this assumption,

\begin{equation}
\langle \delta\theta_{\rm 40K}^2 \rangle
=
\langle \delta\theta_{\rm 130K}^2 \rangle
+
\langle \delta\theta_{\rm add}^2 \rangle
\end{equation}

\noindent which follows from the independence assumption. Consequently, the difference in the order parameter, analogous to Eq.~(9) in the main text, can be written as

\begin{eqnarray}
\Delta S_{\rm fast}^{\rm (40K)} &\simeq& 6\left(
\langle \delta\theta_{\rm 40K}^2 \rangle
-
\langle \delta\theta_{\rm 130K}^2 \rangle
\right) \nonumber \\
&=&
6\,\langle \delta\theta_{\rm add}^2 \rangle,
\label{DS1}
\end{eqnarray}

\noindent where $\Delta S_{\rm fast}^{\rm (40K)}$ quantifies the increase in angular fluctuations relative to the reference state. Although the prefactor is of order unity, the magnitude of $\Delta S_{\rm fast}^{\rm (40K)}$ is determined entirely by $\langle \delta\theta_{\rm add}^2 \rangle$.

From the experimentally obtained fast-axis histograms (Fig.~S8), the standard deviations of the angular distribution are estimated as $\sigma_{\rm 40K}=2.71^\circ=4.73\times10^{-2}~{\rm rad}$ and $\sigma_{\rm 130K}=2.48^\circ=4.33\times10^{-2}~{\rm rad}$. The increase in the mean-square angular fluctuation relative to the high-$T$ reference is then

\begin{eqnarray}
\langle \delta\theta_{\rm add}^2 \rangle &=& \sigma_{\rm 40K}^2 - \sigma_{\rm 130K}^2 \nonumber \\
&\simeq& 3.59\times10^{-4}.
\end{eqnarray}

\noindent Substituting this value into Eq.~(\ref{DS1}) gives

\begin{equation}
\Delta S_{\rm fast}^{\rm (40K)} \simeq 6\,\langle \delta\theta_{\rm add}^2 \rangle \simeq 2.16 \times 10^{-3}.
\end{equation}

\noindent This estimate agrees reasonably well with the observed magnitude of $\Delta S_{10}^{(42\mathrm{K\mbox{-}wnd})}$ in Fig.~3(b), indicating that the characteristic scale of axis-order variation is set by the small angular dispersion of the optical director. Although the fast-axis histogram is analyzed using linear statistics, the angular spread of only a few degrees justifies this approximation. This approach relies on the same small-angle assumption ($\delta\theta\ll1$) used in the previous expansion and is therefore well justified for the present dataset.

\vspace{1cm}

\section*{S4.~Spatial statistical analysis of holonomy variations}

\vspace{0.5cm}

To quantify $T$-induced changes in the holonomy maps, we analyze the variation field $\Delta\omega_{L}^{(5W)}(x,y)$ relative to the lowest-$T$ reference window. For each 5-K-wide window, a representative holonomy map $\omega_{L}^{(5W)}(x,y)$ is computed. Using the lowest-$T$ window (14K-wnd) as reference, the holonomy variation is defined as

\begin{equation}
\Delta \omega_{L}^{(5W)}(x,y)
:=
\omega_{L}^{(5W)}(x,y) - \omega_{L}^{(14\mathrm{K\mbox{-}wnd})}(x,y).
\end{equation}

\noindent Spatial statistical analyses are then performed to characterize the structure of $\Delta\omega_{L}^{(5W)}(x,y)$, including correlation with the director gradient field. In particular, the spatial correlation between $\Delta \omega_{L}^{(5W)}(x,y)$ and $\overline{{\rm grad}}^{(5W)}(x,y)$ is evaluated. Figure~S9(a) shows the $T$ dependence of the Pearson and Spearman correlation coefficients, $\mathrm{corr}(\Delta \omega_{10}^{(5W)}, \,\overline{{\rm grad}}^{(5W)})$, computed over all valid pixels $(x,y)$ in the field of view. The coefficients are approximately $0.5$--$0.6$, indicating moderate correlation between the holonomy variation and local-gradient magnitude. This result demonstrates that the spatial variations of $\Delta \omega_{10}^{(5W)}$ are not solely determined by the magnitude of the local director gradient, but also reflect orientational incompatibilities accumulated along closed loops in the director field.

\begin{figure}[tb] 
\begin{center} 
\includegraphics[width=15cm]{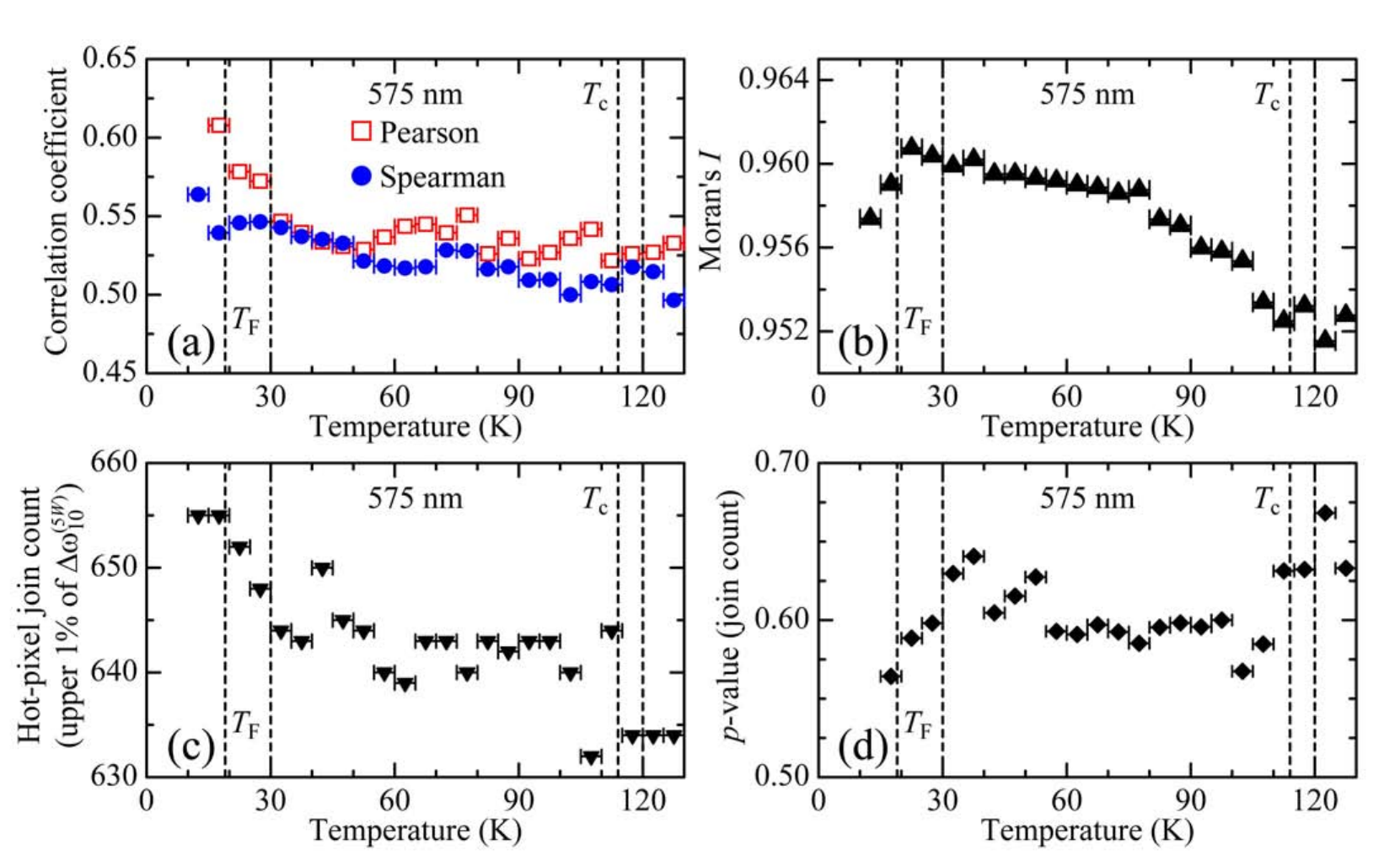} 
\end{center} 
\caption{Spatial statistical analysis of holonomy variations. (a) Temperature dependence of the correlation between $\Delta \omega_{10}^{(5W)}(x,y)$ and $\overline{{\rm grad}}^{(5W)}(x,y)$, evaluated over all valid pixels. Pearson (open squares) and Spearman (closed circles) correlation coefficients are shown. (b) Moran's $I$ of the $\Delta\omega_{10}^{(5W)}(x,y)$ field, quantifying spatial autocorrelation, evaluated over all valid pixels. (c) Join count of ``hot pixels,'' defined as the upper 1\% of $\Delta\omega_{10}^{(5W)}$ values, measuring the number of adjacent hot-pixel pairs. (d) $p$-values from a torus-shift permutation test (10,000 random realizations) for the join-count statistic. Dashed vertical lines indicate the reported ranges of $T_{\rm c}$ and $T_{\rm F}$.}
\end{figure} 

We next evaluate Moran's $I$, a widely used measure of spatial autocorrelation, to quantify the spatial organization of the $\Delta\omega_{10}^{(5W)}$ field over all valid pixels. Moran's $I$ is defined as

\begin{equation}
I =
\frac{ \mathcal{N}_{10}\sum_{(x,y)}\sum_{(x',y')} 
c_{(x,y),(x',y')} 
\left(\Delta\omega_{10}^{(5W)}(x,y)-\langle \Delta\omega_{10}^{(5W)} \rangle_{xy} \right)
\left(\Delta\omega_{10}^{(5W)}(x',y')-\langle \Delta\omega_{10}^{(5W)} \rangle_{xy} \right) }
{ \left(\sum_{(x,y)}\sum_{(x',y')} c_{(x,y),(x',y')}\right)
\left( \sum_{(x,y)} 
\left(\Delta\omega_{10}^{(5W)}(x,y)-\langle \Delta\omega_{10}^{(5W)} \rangle_{xy}\right)^2\right) } .
\end{equation}

\noindent where $\mathcal{N}_{10}$ is the total number of valid pixels for a $10\times10$ loop and $\langle \Delta\omega_{10}^{(5W)} \rangle_{xy}$ is the spatial mean,

\begin{equation}
\langle \Delta\omega_{10}^{(5W)} \rangle_{xy} = \frac{\sum_{(x,y)} \Delta\omega_{10}^{(5W)}(x,y)}{\mathcal{N}_{10}}.
\end{equation}

\noindent The adjacency weights $c_{(x,y),(x',y')}$ are defined as

\begin{equation}
c_{(x,y),(x',y')} =
\begin{cases}
1 & \text{if }\, |x-x'|+|y-y'|=1,\\
0 & \text{otherwise},
\end{cases}
\end{equation}

\noindent with $c_{(x,y),(x,y)}=0$.

Figure~S9(b) shows the $T$ dependence of Moran's $I$. The values are relatively large ($I \simeq 0.95$--0.96), indicating strong spatial autocorrelation in the $\Delta\omega_{10}^{(5W)}$ field. That is, holonomy variations form spatially coherent structures rather than random noise. The high Moran's $I$ partly reflects the strongly skewed distribution of $\Delta\omega_{10}^{(5W)}(x,y)$, where most pixels satisfy $\Delta\omega_{10}^{(5W)} \simeq 0$. Neighboring pixels therefore frequently share similarly small values, making the covariance term in Moran's $I$ predominantly positive and naturally inflating $I$. Since the analysis is performed on the difference map $\Delta\omega_{10}^{(5W)}$ relative to the 14K-wnd, static background structures are largely removed. Consequently, the large Moran's $I$ values mainly reflect spatial correlations in the holonomy variations themselves, rather than only static background structures. This conclusion, based on statistics over all pixels, is not limited to regions of high $\Delta\omega_{10}^{(5W)}$.

To examine the spatial clustering of large holonomy variations, pixels in the upper 1\% of $\Delta\omega_{10}^{(5W)}$ are defined as ``hot pixels.'' Because of $\mathcal{N}_{10} = (302-10+1)\times(140-10+1)=38{,}383$, the upper 1\% corresponds to approximately 384 pixels. Figure~S9(c) shows the join count of hot pixels, which measures the number of adjacent hot-pixel pairs and quantifies the degree of spatial clustering among extreme values. The observed join counts are roughly 630--660 with little $T$ dependence. This is much larger than the random expectation ($\simeq 8$ for a hot-pixel fraction $p=0.01$ on a 4-neighbor lattice assuming independent Bernoulli statistics) but still below the value for a highly compact connected cluster of comparable size. For example, a rectangular $16\times24$ cluster containing 384 pixels has a join count of $728 \,(=384\times2-16-24)$. These results indicate that extreme holonomy variations exhibit substantial spatial aggregation but do not form maximally compact clusters; instead, hot pixels are distributed across spatially extended, non-compact regions.

To assess the statistical significance of the spatial structure, we perform a permutation test using a torus-shift procedure as the null model. In this procedure, the $\Delta\omega_{10}^{(5W)}$ map is randomly translated with periodic boundary conditions, which preserves the value distribution and local spatial autocorrelation while randomizing the global configuration. For each $T$ window, 10,000 random realizations are generated, and the $p$-value is calculated as

\begin{equation} 
p = P(\mathrm{Stat}_{\mathrm{random}} \ge \mathrm{Stat}_{\mathrm{observed}}) \qquad \text{(one-sided test)}.
\end{equation}

\noindent Figure~S9(d) shows the $p$-values for the join-count statistic. We find $p \simeq 0.55$--0.70, indicating that the observed clustering of hot pixels is not significantly different from the null model. In contrast, the permutation test for Moran's $I$ yields $p < 10^{-4}$ for all $T$ windows, confirming statistically significant spatial autocorrelation in the $\Delta\omega_{10}^{(5W)}$ field. These results demonstrate that holonomy variations are strongly spatially correlated but do not form tightly localized clusters of extreme values. This conclusion is consistent with three observations: (i) holonomy shows only moderate correlation with the gradient magnitude, (ii) it exhibits significant spatial autocorrelation, and (iii) extreme values are spatially extended rather than sharply localized.

\begin{figure}[tb] 
\begin{center} 
\includegraphics[width=15cm]{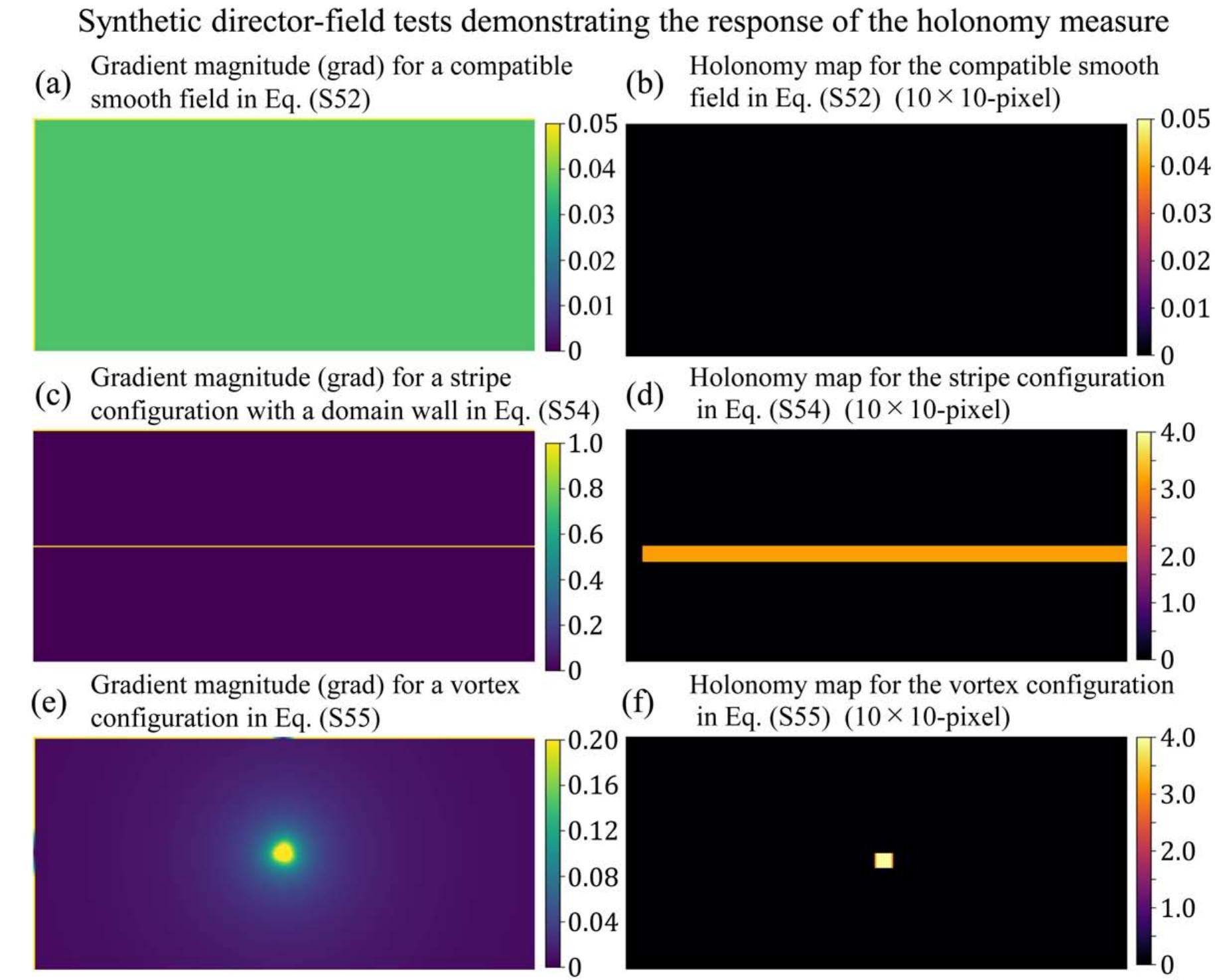} 
\end{center} 
\caption{Synthetic tests illustrating the response of the holonomy measure to different director-field configurations (grad). (a,b) Smooth compatible field $\theta(x,y)=0.03x+0.02y$: (a) ${\rm grad}(x,y)$ and (b) $\omega_{10}(x,y)$ for a $10\times10$-pixel loop. (c,d) Stripe configuration with a domain wall ($\theta=0 \to \pi/2$): (c) grad and (d) holonomy. (e,f) Vortex configuration $\theta(x,y)=\mathrm{atan2}(y-y_c,x-x_c)$ representing a topological defect: (e) grad and (f) holonomy.}
\end{figure}

To verify that the holonomy measure captures incompatible rotational structures rather than simply the magnitude of the local gradient, we perform numerical tests using simple synthetic director-field configurations, $\mathbb{R}P^2$. Figure~S10 shows ${\rm grad}(x,y)$ and $\omega_{10}(x,y)$ for three representative configurations. First, we consider a smooth, compatible field

\begin{equation}
\theta(x,y)=0.03x+0.02y,
\end{equation}

\noindent where $\theta(x,y)$ denotes the local director orientation and $\nabla\theta(x,y)$ is the corresponding gradient field. In this case, the gradient is spatially uniform and nonzero. Because the field is compatible, the rotation accumulated along any closed loop cancels exactly:

\begin{equation}
\oint \nabla \theta(x,y) \cdot dl = 0.
\end{equation}

\noindent This behavior follows from the fact that the gradient field is conservative: a smooth, compatible gradient field does not produce rotational incompatibility, and thus the holonomy around closed loops vanishes (up to numerical errors). As shown in Figs.~S10(a) and S10(b), the grad map exhibits finite values, whereas the holonomy map is nearly zero. Next, we consider a stripe structure with a domain wall at the center, across which the director orientation changes discontinuously:

\begin{equation} 
\theta = 0 \quad \rightarrow \quad \theta = \pi/2.
\end{equation}

\noindent Here, the grad map shows large values along the interface (Fig.~S10(c)), and the holonomy map detects the discontinuity, displaying finite values wherever $10\times10$-pixel loops cross the domain wall (Fig.~S10(d)). Finally, we examine a vortex-like configuration representing a topological defect:

\begin{equation} 
\theta(x,y)=\mathrm{atan2}\,(y-y_c, x-x_c), 
\end{equation}

\noindent with the defect core at $(x_c,y_c)=(151,70)$, the center of the $302\times140$ grid. In this case, the gradient is large near the defect core (Fig.~S10(e)), and the holonomy map exhibits a strong, localized peak around the defect, reflecting the winding structure of the director field (Fig.~S10(f)). 

These synthetic tests highlight the qualitative distinction between the gradient- and holonomy-based measures. While the gradient magnitude reflects local angular variations of the director field, the holonomy specifically captures rotational incompatibility accumulated along closed loops. As a result, smooth compatible fields yield vanishing holonomy despite finite gradients, whereas domain walls and topological defects generate localized holonomy signals. This demonstrates that the holonomy measure is sensitive to the geometrical incompatibilities of the director field, rather than being determined solely by large local gradients.

\vspace{1cm}

\section*{S5.~Real-space comparison of the axis-ordering change}

\vspace{0.5cm}

To examine the evolution of spatial axis-ordering patterns upon cooling, we compare each lower-$T$ 5-K-wide window with a high-$T$ reference state, chosen as the 132K-wnd. Let $S_{L}^{(T)}(x,y)$ denote the pixel-wise axis-order map at $T$. The representative map $\overline{S}_{L}^{(5W)}(x,y)$ and its variation $\Delta S_L^{(5W)}(x,y)$ are defined as in Eqs.~(8) and (9) of the main text. While the global analysis in Sec.~S3 shows that $\Delta S_{\rm fast}^{\rm (40K)}$ reflects the increase in angular fluctuations of the fast-axis orientation upon cooling and is therefore positive definite, the local quantity $\Delta S_{L}^{(5W)}(x,y)$ represents spatial deviations relative to the reference state and can take both positive and negative values.

In the present analysis, the local estimate of $S_{L}^{(T)}(x,y)$ becomes unstable when the corresponding holonomy angle is extremely small, because the associated rotation axis is then weakly constrained. To avoid overinterpreting such poorly defined regions, we introduce a common quality filter based on the magnitude of $\omega_{L}^{(T)}$. Specifically, we pool all valid representative holonomy angles $\overline{\omega}_{L}^{(5W)}(x,y)$ across all windows up to and including the 132K-wnd, and define a single global threshold $\omega_{L, {\rm th}}$ as the 25th percentile of the pooled distribution:

\begin{equation}
\omega_{L, {\rm th}} := {\rm percentile}_{25} \left( \bigcup_{5W\le 132\mathrm{K\mbox{-}wnd}} \left\{ \overline{\omega}_{L}^{(5W)}(x,y) \;\big|\; (x,y)\in \Omega_{L, {\rm all}}^{(5W)} \right\} \right), \label{eq:S3_omega_th} 
\end{equation}

\noindent where $\Omega_{L, {\rm all}}^{(5W)}$ denotes the set of all valid pixels for an $L\times L$ loop in the corresponding 5-K-wide window. Including the 132K-wnd in the pooled distribution ensures that the quality filter is defined consistently over the full set of windows used in the $\Delta S_{L}^{(5W)}(x,y)$ comparison, including the high-$T$ reference. For each window, we then define the subset of pixels with holonomy angles below this threshold as

\begin{equation} 
\Omega_{L, 25}^{(5W)} := \left\{ (x,y)\in\Omega_{L, {\rm all}}^{(5W)} \;\big|\; \overline{\omega}_{L}^{(5W)}(x,y)<\omega_{L, {\rm th}} \right\}. 
\end{equation} 

\noindent We define the excluded set as the union over all windows included in the $\Delta S_{L}^{(5W)}$ analysis:

\begin{equation} 
\Omega_{L, {\rm excl}}^{(5W)} := \bigcup_{5W\le 132\mathrm{K\mbox{-}wnd}} \Omega_{L, 25}^{(5W)}, 
\end{equation} 

\noindent and the common valid pixel set is then

\begin{equation} 
\Omega_{L, {\rm valid}}^{(5W)} := \Omega_{L, {\rm all}}^{(5W)}\setminus \Omega_{L, {\rm excl}}^{(5W)}. \label{eq:S3_valid} 
\end{equation}

\noindent Pixels outside $\Omega_{L, {\rm valid}}^{(5W)}$ are displayed transparently in the visualizations. This construction ensures that all $\Delta S_{L}^{(5W)}(x,y)$ maps are compared over a consistent, reliable pixel set and are not affected by the window-dependent disappearance of the axis signal.

As shown in Figs.~S11--S16, the $\Delta S_{10}^{(5W)}(x,y)$ maps reveal that the spatial distribution of axis-ordering changes varies substantially near both $T_{\rm c}$ and $T_{\rm F}$. While the maximum values are comparable between the 14K-wnd and 42K-wnd, their spatial patterns differ markedly. The $\Delta S_{10}^{(5W)}$ pattern for the 42K-wnd closely resembles the grad map in Fig.~1(c), consistent with a strain-gradient-related origin of the underlying incompatibility structure. In contrast, the pattern for the 14K-wnd differs from both the low-$T$ holonomy map and the grad map shown in Fig.~S3(a). This indicates that the low-$T$ reorganization of the axis-ordering field cannot be fully explained by strain or stress concentration alone. Instead, additional polarization-related mechanisms---likely associated with ferroelectric-domain formation and domain configurations---appear to contribute to shaping the real-space structure of the holonomy response.

\begin{figure}[b] 
\begin{center} 
\includegraphics[width=12cm]{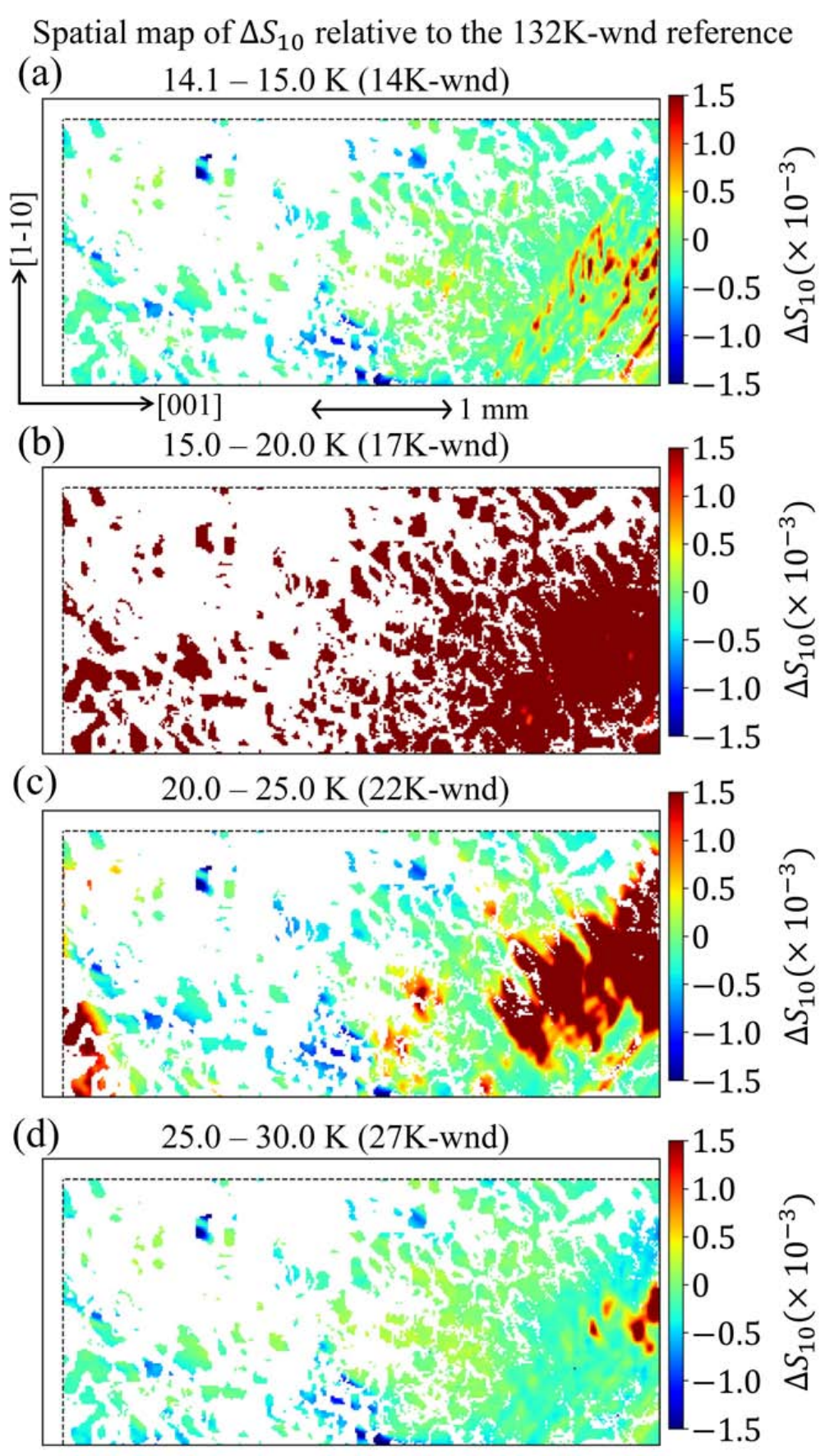} 
\end{center} 
\caption{Maps of $\Delta S_{10}^{(5W)}(x,y)$ for (a) 14K-wnd, (b) 17K-wnd, (c) 22K-wnd, and (d) 27K-wnd. As in Fig.~S1, regions above and to the left of the dashed lines are excluded from the analysis to avoid boundary artifacts.} 
\end{figure}

\clearpage 

\begin{figure}[b] 
\begin{center} 
\includegraphics[width=12cm]{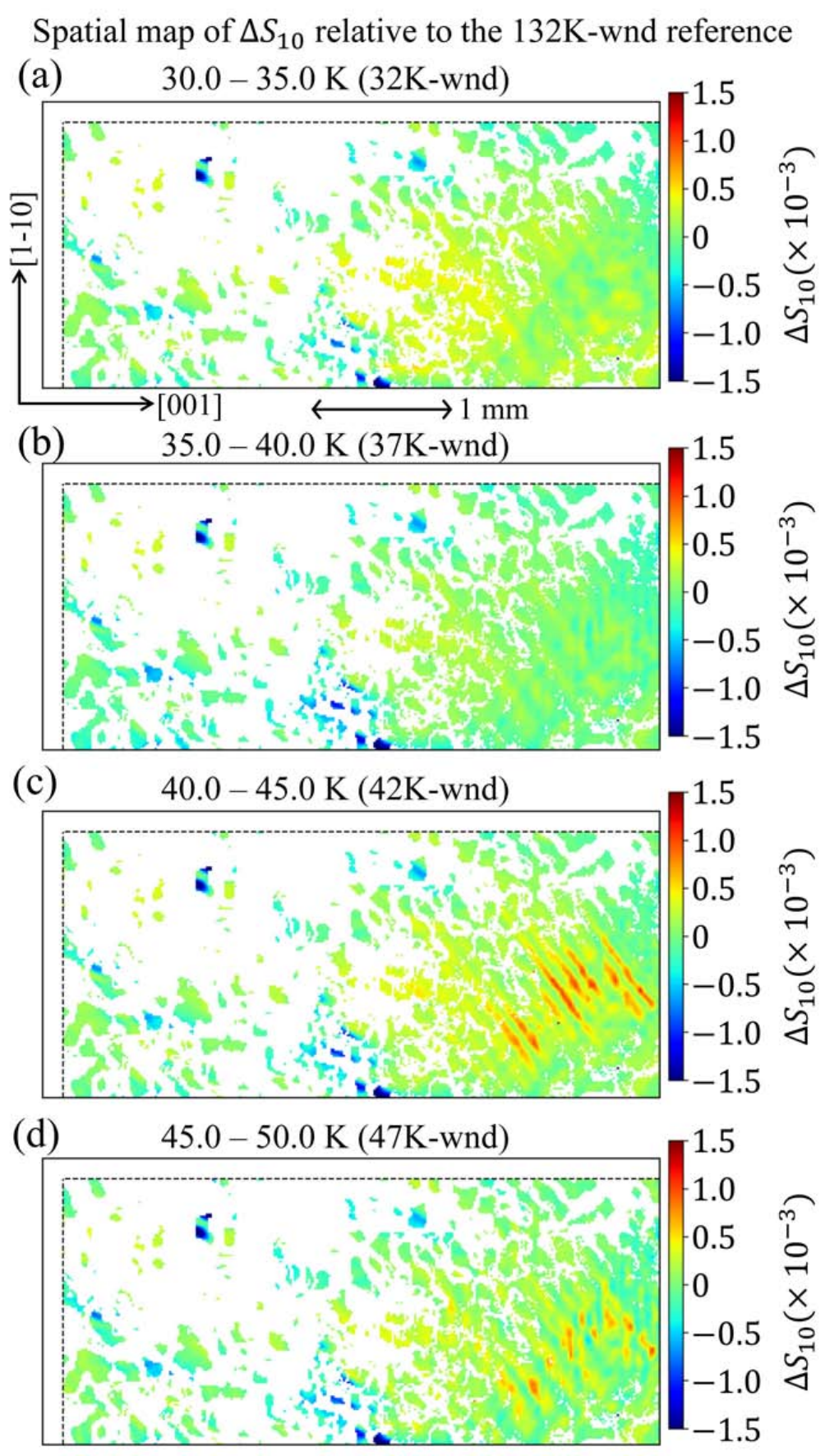} 
\end{center} 
\caption{Maps of $\Delta S_{10}^{(5W)}(x,y)$ for (a) 32K-wnd, (b) 37K-wnd, (c) 42K-wnd, and (d) 47K-wnd. As in Fig.~S1, regions above and to the left of the dashed lines are excluded from the analysis to avoid boundary artifacts.} 
\end{figure} 

\clearpage 

\begin{figure}[b] 
\begin{center} 
\includegraphics[width=12cm]{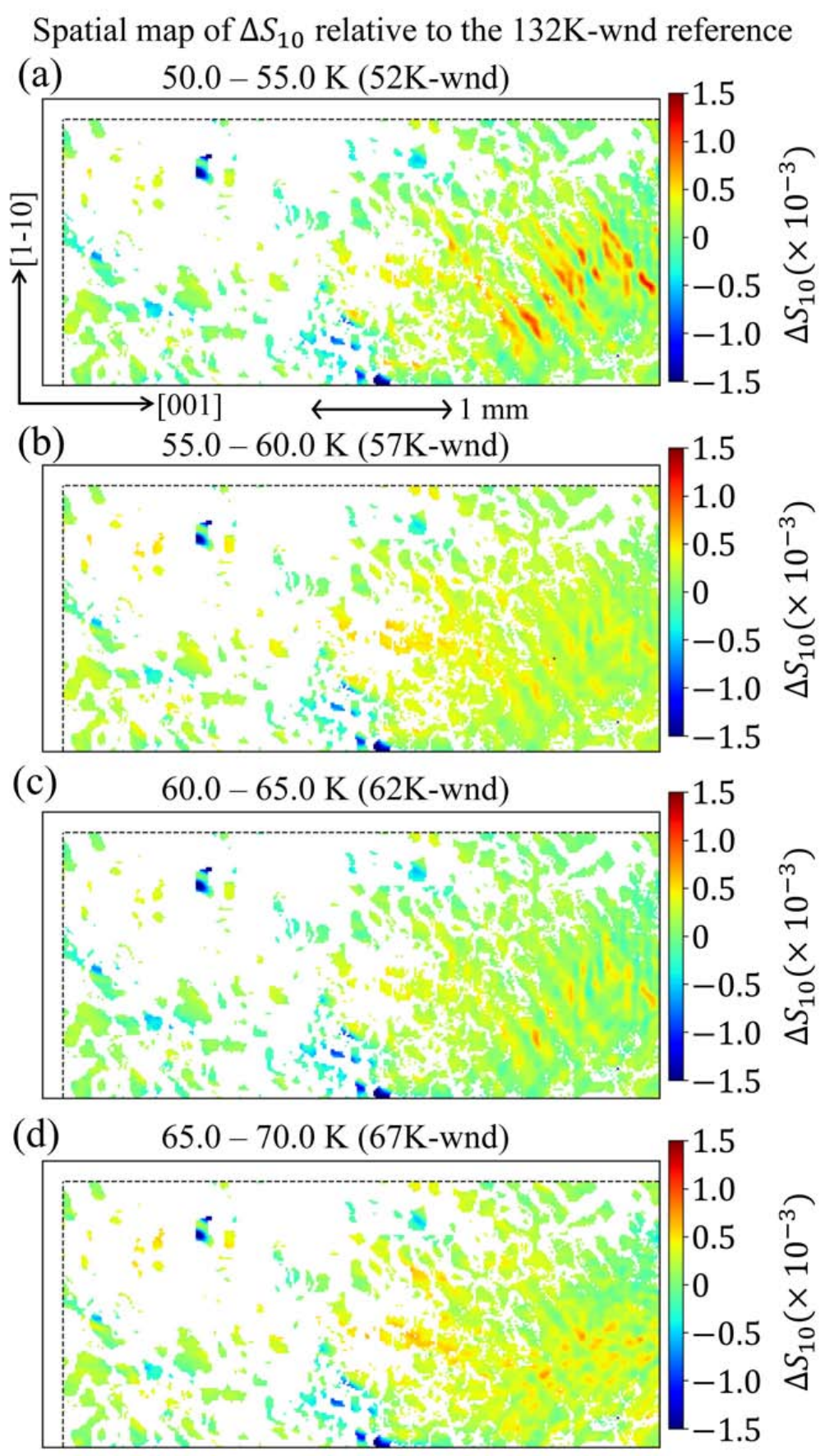} 
\end{center} 
\caption{Maps of $\Delta S_{10}^{(5W)}(x,y)$ for (a) 52K-wnd, (b) 57K-wnd, (c) 62K-wnd, and (d) 67K-wnd. As in Fig.~S1, regions above and to the left of the dashed lines are excluded from the analysis to avoid boundary artifacts.} 
\end{figure} 

\clearpage 

\begin{figure}[b] 
\begin{center} 
\includegraphics[width=12cm]{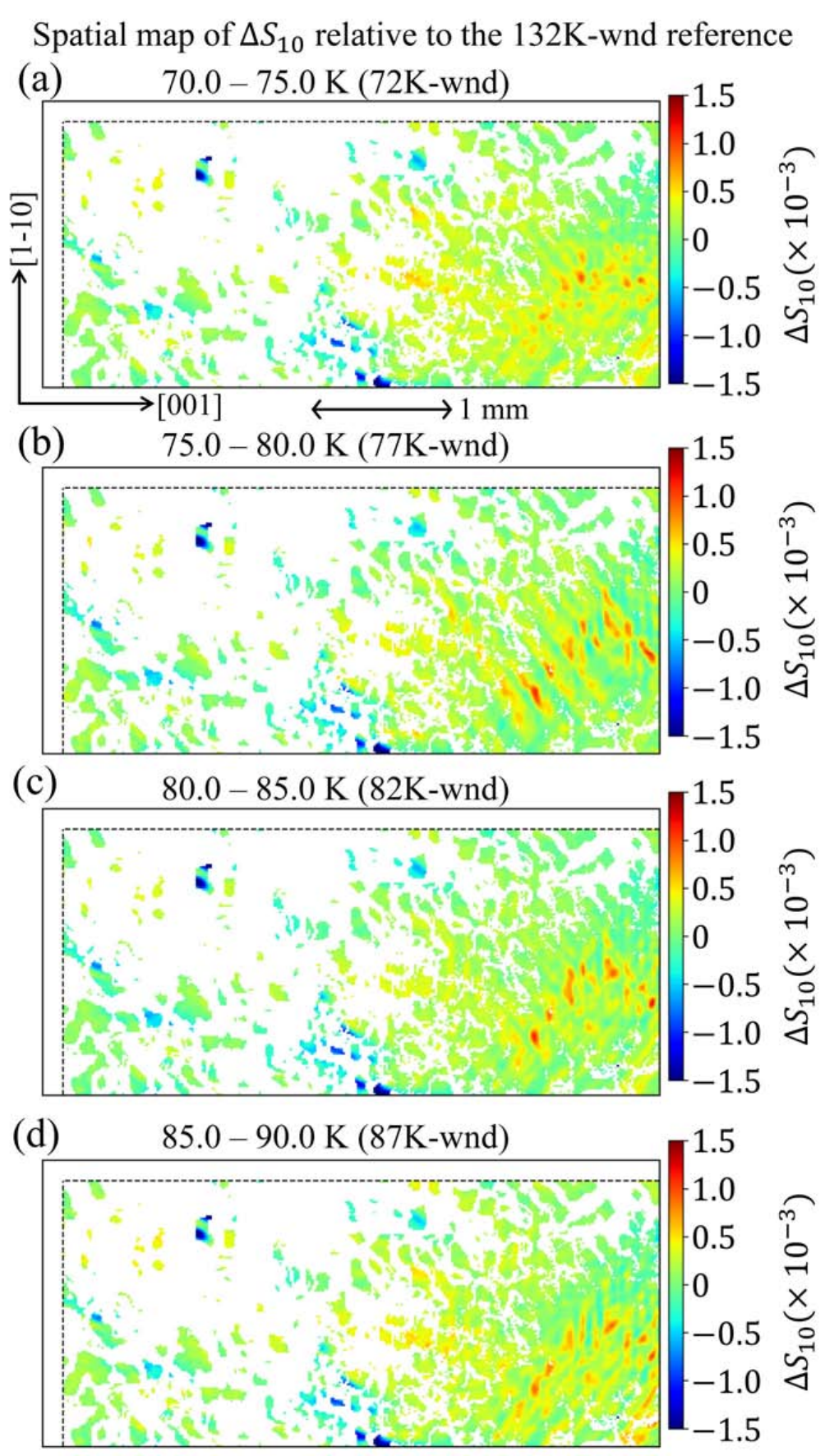} 
\end{center} 
\caption{Maps of $\Delta S_{10}^{(5W)}(x,y)$ for (a) 72K-wnd, (b) 77K-wnd, (c) 82K-wnd, and (d) 87K-wnd. As in Fig.~S1, regions above and to the left of the dashed lines are excluded from the analysis to avoid boundary artifacts.} 
\end{figure} 

\clearpage 

\begin{figure}[b] 
\begin{center} 
\includegraphics[width=12cm]{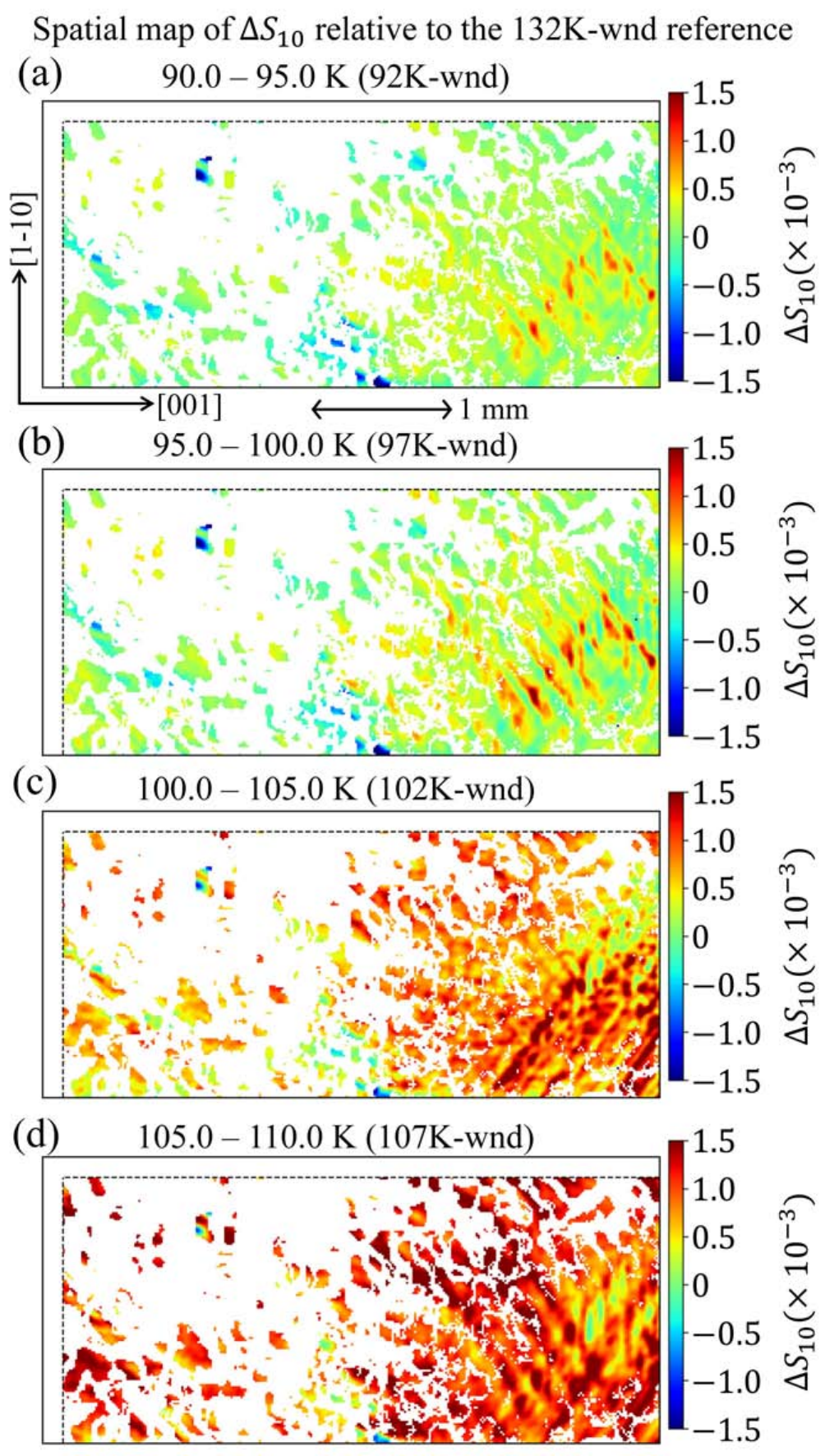} 
\end{center} 
\caption{Maps of $\Delta S_{10}^{(5W)}(x,y)$ for (a) 92K-wnd, (b) 97K-wnd, (c) 102K-wnd, and (d) 107K-wnd. As in Fig.~S1, regions above and to the left of the dashed lines are excluded from the analysis to avoid boundary artifacts.} 
\end{figure} 

\clearpage 

\begin{figure}[b] 
\begin{center} 
\includegraphics[width=12cm]{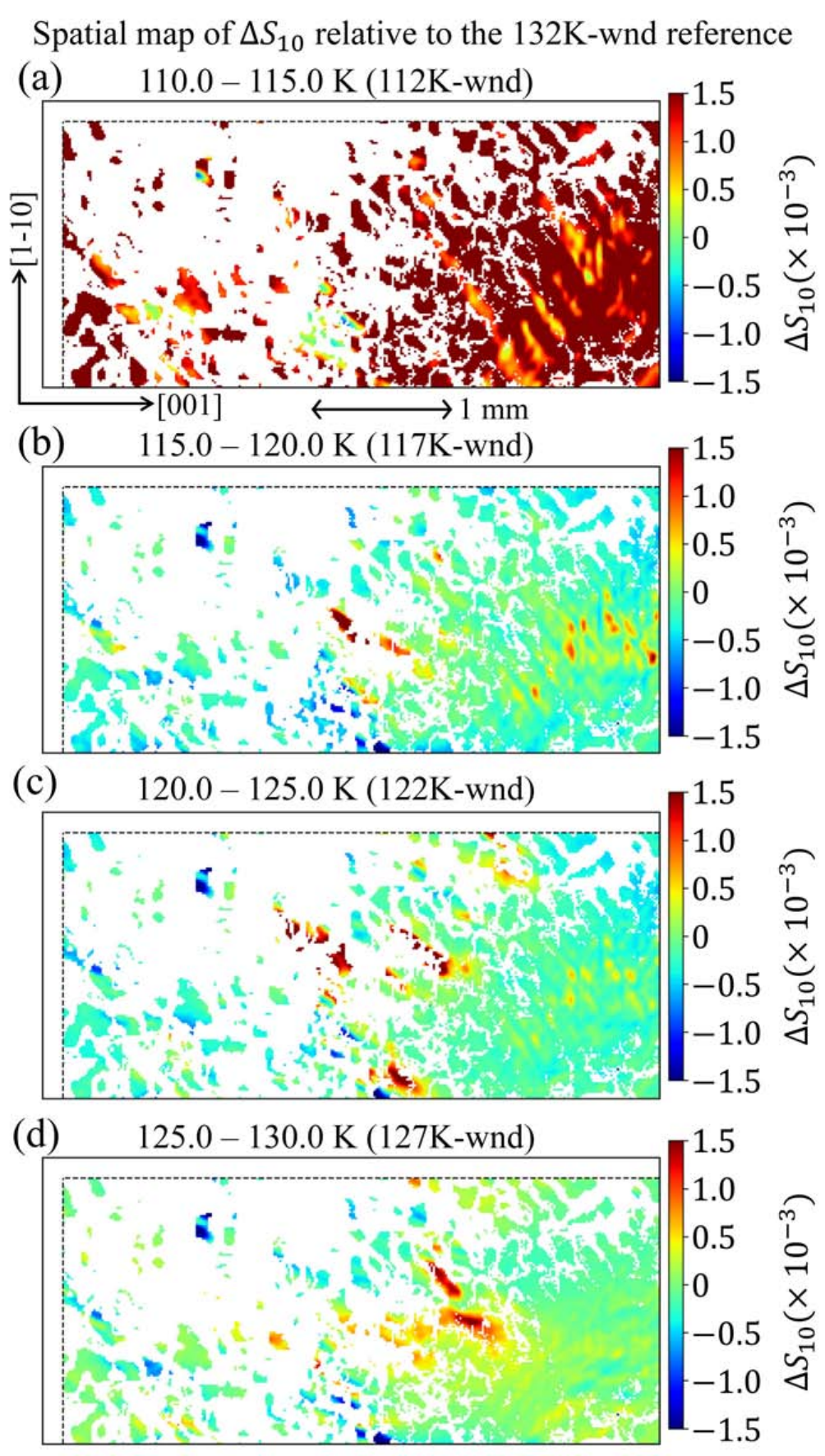} 
\end{center} 
\caption{Maps of $\Delta S_{10}^{(5W)}(x,y)$ for (a) 112K-wnd, (b) 117K-wnd, (c) 122K-wnd, and (d) 127K-wnd. As in Fig.~S1, regions above and to the left of the dashed lines are excluded from the analysis to avoid boundary artifacts.} 
\end{figure}

\clearpage

\section*{S6.~Quantification of inter-window reorganization}

\vspace{0.5cm}

To quantify the degree of real-space reorganization of $\Delta S_{L}^{(5W)}(x,y)$ between adjacent $T$ windows, we compare consecutive 5-K-wide windows in descending order of $T$. Let $W_j$ denote the higher-$T$ window (base) and $W_{j+1}$ the immediately lower-$T$ window (current). Because $\Delta S_{L}^{(5W)}$ is meaningful only where the holonomy-axis direction is reliably defined, the analysis is first restricted to the valid pixel set of the base window $\Omega_{L, {\rm valid}}^{(5W_j)}$. Within this region, we further focus on pixels exhibiting the strongest holonomy response in the base window. Specifically, we define the upper 10\% region of $|\Delta S_{L}^{(5W_j)}(x,y)|$ within the valid set as

\begin{eqnarray}
\Omega_{L, 10\%}^{(5W_j)}
&:=&
\left\{
(x,y)\in\Omega_{L, {\rm valid}}^{(5W_j)}
\;|\;
|\Delta S_{L}^{(5W_j)}(x,y)|
\ge
\tau_{L, 0.9}^{(5W_j)}
\right\},
\label{eq:S4_top10} \\
\tau_{L, 0.9}^{(5W_j)}
&:=&
{\rm percentile}_{90}
\left(
\left\{
|\Delta S_{L}^{(5W_j)}(x,y)|
\;\big|\;
(x,y)\in \Omega_{L, {\rm valid}}^{(5W_j)}
\right\}
\right).\label{eq:S4_top10-2}
\end{eqnarray}

\noindent For each adjacent window pair, the pixels in $\Omega_{L, 10\%}^{(5W_j)}$ are indexed as

\begin{equation}
(x_i,y_i)\in\Omega_{L, 10\%}^{(5W_j)}
\qquad
i = 1,\ldots,\mathcal{N}_{L}^j
\qquad
\mathcal{N}_{L}^j := \left|\Omega_{L, 10\%}^{(5W_j)}\right|.
\label{eq:S4_pixels}
\end{equation}

\noindent The two spatial fields are then restricted to this set:

\begin{eqnarray}
X_i &:=& \Delta S_{L}^{(5W_j)}(x_i,y_i), \\
Y_i &:=& \Delta S_{L}^{(5W_{j+1})}(x_i,y_i).
\label{eq:S4_vectorization}
\end{eqnarray}

\noindent Four complementary indices are evaluated for each adjacent pair. First, the Pearson correlation coefficient is 

\begin{equation} 
{\rm corr}_j := \frac{ \sum_{i=1}^{\mathcal{N}_{L}^j}(X_i-\overline{X})(Y_i-\overline{Y}) }{ \sqrt{\sum_{i=1}^{\mathcal{N}_{L}^j}(X_i-\overline{X})^2}\, \sqrt{\sum_{i=1}^{\mathcal{N}_{L}^j}(Y_i-\overline{Y})^2} }, \label{eq:S4_corr} 
\end{equation} 

\noindent where $\overline{X}$ and $\overline{Y}$ denote the means of $\{X_i\}$ and $\{Y_i\}$, respectively. Second, we examine whether the current pattern can be approximated by an affine transformation of the base pattern: 

\begin{equation} 
Y_i \simeq a_j X_i + b_j \qquad i=1,\ldots,\mathcal{N}_{L}^j, 
\end{equation} 

\noindent where $a_j$ and $b_j$ are the slope and intercept, estimated via ordinary least squares. From this fit, the coefficient of determination is

\begin{equation} 
R_j^2 := 1- \frac{ \sum_{i=1}^{\mathcal{N}_{L}^j} \bigl(Y_i-(a_jX_i+b_j)\bigr)^2 }{ \sum_{i=1}^{\mathcal{N}_{L}^j}(Y_i-\overline{Y})^2 }. \label{eq:S4_R2} 
\end{equation} 

\noindent Third, the fit residual is quantified using the normalized root mean square error (NRMSE): 

\begin{eqnarray} 
{\rm RMSE}_j &:=& \sqrt{ \frac{1}{\mathcal{N}_{L}^j} \sum_{i=1}^{\mathcal{N}_{L}^j} \bigl(Y_i-(a_jX_i+b_j)\bigr)^2 }, \\ 
{\rm NRMSE}_j &:=& \frac{{\rm RMSE}_j}{\sigma_Y+\varepsilon}, \label{eq:S4_nrmse} 
\end{eqnarray} 

\noindent where $\sigma_Y$ is the sample standard deviation of $\{Y_i\}$ and $\varepsilon=1\times 10^{-12}$ is introduced to prevent division by zero. Fourth, we compute the sign-flip rate:

\begin{equation} 
f_{{\rm flip},j} := \frac{1}{\mathcal{N}_{L}^j} \left| \left\{ i\, \big|\; X_iY_i<0 \right\} \right|. \label{eq:S4_flip} 
\end{equation} 

\noindent This metric represents the fraction of pixels for which the local change reverses sign between adjacent windows. Because the four indices have different numerical scales, each metric is standardized across all adjacent window pairs. To ensure that larger values consistently correspond to stronger pattern changes, similarity-based metrics (corr and $R^2$) are converted to dissimilarity measures using $(1-\mathrm{corr})$ and $(1-R^2)$ prior to standardization. Specifically, we define

\begin{eqnarray} 
Z_j^{(1-{\rm corr})} &:=& \frac{(1-{\rm corr}_j)-\mu_{(1-{\rm corr})}}{\sigma_{(1-{\rm corr})}} ,\\ 
Z_j^{(1-R^2)} &:=& \frac{(1-R_j^2)-\mu_{(1-R^2)}}{\sigma_{(1-R^2)}}, \\ 
Z_j^{({\rm NRMSE})} &:=& \frac{{\rm NRMSE}_j-\mu_{\rm NRMSE}}{\sigma_{\rm NRMSE}}, \\
Z_j^{(f_{\rm flip})} &:=& \frac{f_{{\rm flip},j}-\mu_{f_{\rm flip}}}{\sigma_{f_{\rm flip}}},  \label{CS}
\end{eqnarray}

\noindent where each $\mu$ and $\sigma$ denotes the mean and standard deviation of the corresponding metric over all adjacent window pairs. The overall degree of inter-window reorganization is then summarized by the unweighted mean of the four standardized metrics:

\begin{figure}[b] 
\begin{center} 
\includegraphics[width=15cm]{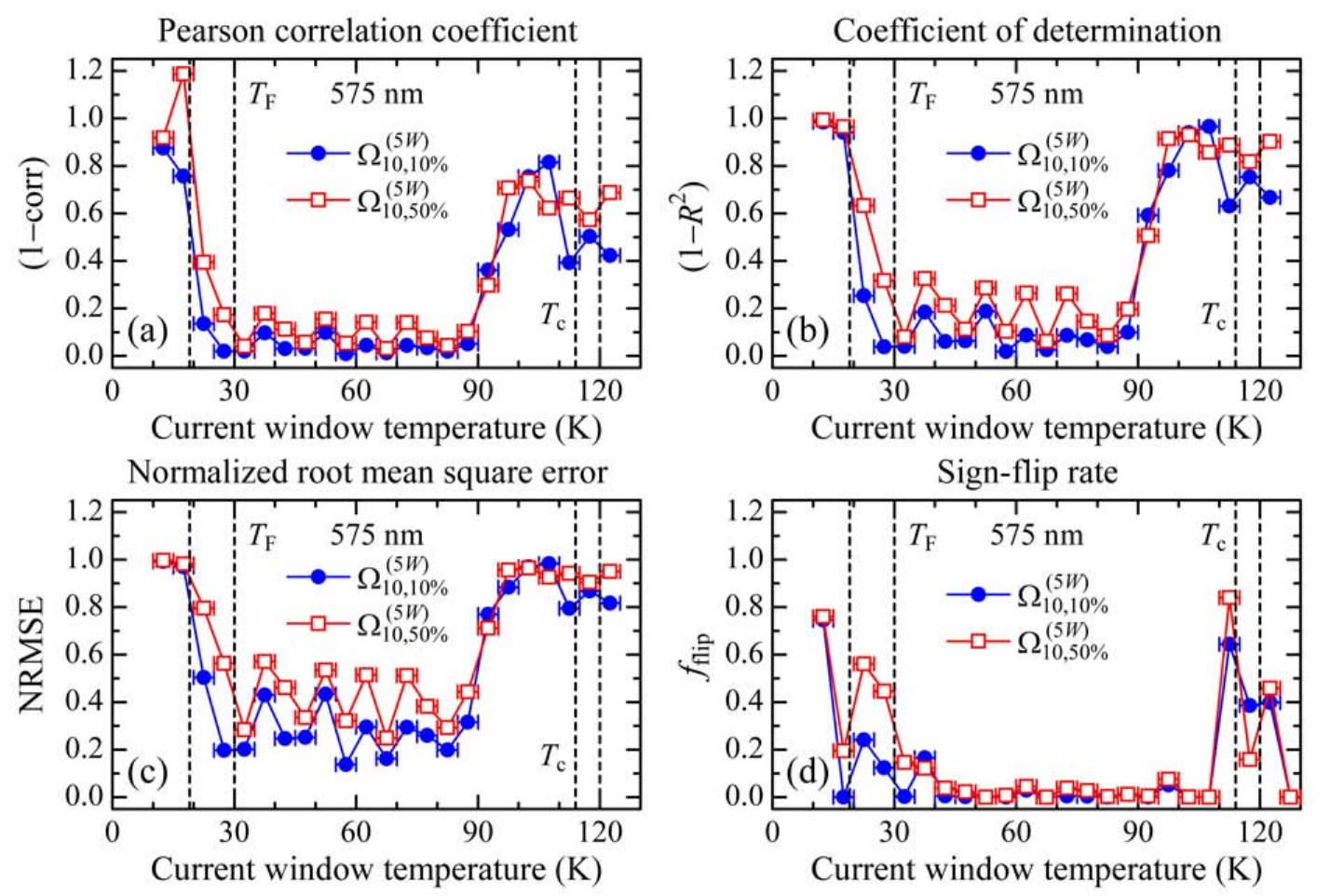} 
\end{center} 
\caption{Temperature dependence of statistical metrics characterizing the reorganization of the real-space pattern of $\Delta S_{L}^{(5W)}(x,y)$ between adjacent $T$ windows: (a) Pearson correlation coefficient $(1-{\rm corr})$, (b) coefficient of determination $(1-R^2)$, (c) normalized root mean square error (NRMSE), and (d) sign-flip rate $f_{\rm flip}$. The horizontal axis denotes the temperature of the current window ($5W_{j+1}$) in each adjacent pair $(5W_j,5W_{j+1})$. Dashed vertical lines indicate the reported ranges of $T_{\rm c}$ and $T_{\rm F}$.}
\end{figure} 

\begin{figure}[tb] 
\begin{center} 
\includegraphics[width=10cm]{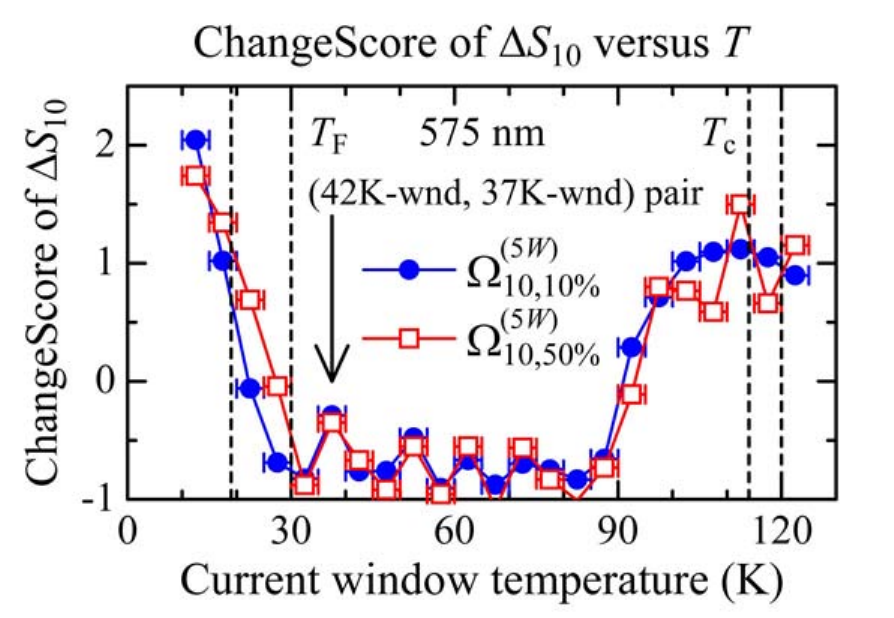} 
\end{center} 
\caption{Temperature dependence of the ${\rm ChangeScore}$ of $\Delta S_{10}^{(5W)}$. The ${\rm ChangeScore}$, defined as the average of the four standardized metrics introduced in Eqs.~(S71)--(S75), quantifies the degree of reorganization of the $\Delta S_{10}^{(5W)}$ pattern between adjacent $T$ windows. The horizontal axis denotes the temperature of the current window ($5W_{j+1}$) in each adjacent pair $(5W_j,5W_{j+1})$. Dashed vertical lines indicate the reported ranges of $T_{\rm c}$ and $T_{\rm F}$.} 
\end{figure}

\begin{equation} 
{\rm ChangeScore}_j := \frac{Z_j^{(1-{\rm corr})} + Z_j^{(1-R^2)} + Z_j^{({\rm NRMSE})} + Z_j^{(f_{\rm flip})}}{4}. \label{eq:S4_change_score}
\end{equation} 

\noindent Larger values of ${\rm ChangeScore}_j$ correspond to stronger reorganization of the $\Delta S_{L}^{(5W)}$ pattern between adjacent windows. 

Figure~S17 shows the $T$ dependence of the four statistical metrics at scale $L=10$: (a) $(1-{\rm corr})$, (b) $(1-R^2)$, (c) NRMSE, and (d) $f_{\rm flip}$. The analysis was also repeated using a larger threshold (the upper 50\% region) by replacing Eqs.~(\ref{eq:S4_top10}) and (\ref{eq:S4_top10-2}) with the corresponding percentile definition. The results exhibit the same qualitative trends, indicating that the analysis is robust to the choice of threshold. These findings suggest that the reorganization of the real-space pattern of $\Delta S_{10}^{(5W)}$ can be broadly classified into three $T$ regimes. Here, the term ``regime'' does not imply distinct thermodynamic phases, but rather $T$ ranges characterized by statistically distinct reorganization behaviors. In the high-$T$ region (approximately 90--120~K), all metrics assume relatively large values, reflecting substantial changes in the spatial pattern of $\Delta S_{10}^{(5W)}$ between adjacent $T$ windows. Both the amplitude and the sign structure vary significantly in the high-$T$ region, suggesting that the orientational structure of the holonomy-axis field is still undergoing reorganization. In the intermediate-$T$ region (approximately 30--80~K), all metrics are relatively small, indicating that the spatial pattern of $\Delta S_{10}^{(5W)}$ is largely preserved between adjacent $T$ windows. Both the amplitude and sign structure remain stable, implying that the orientational texture of the holonomy-axis field is effectively frozen within this range. In the low-$T$ region (below 30~K), all metrics increase again, signaling renewed reorganization of the $\Delta S_{10}^{(5W)}$ pattern accompanied by sign reversals. This behavior reflects a substantial rearrangement of the holonomy-axis structure, potentially associated with a reconfiguration of the orientational texture underlying the low-$T$ electromechanical response. Overall, these results indicate that the $T$ evolution of the holonomy-axis pattern is non-monotonic, proceeding through three statistically distinct regimes.

Figure~S18 shows the $T$ dependence of ${\rm ChangeScore}_j$ as defined in Eq.~(\ref{eq:S4_change_score}). The results exhibit the same qualitative trends, indicating that ${\rm ChangeScore}_j$ is robust to the choice of threshold. The composite score displays pronounced peaks near $T_{\rm c}$ and $T_{\rm F}$, reflecting substantial reorganization of the $\Delta S_{10}^{(5W)}$ pattern in the vicinity of these characteristic temperatures. Away from these transitions, ${\rm ChangeScore}_j$ remains relatively small across most of the $T$ range. A minor secondary feature is observed around the (42K-wnd, 37K-wnd) pair, suggesting a slight additional modification of the spatial pattern within this temperature interval.

For the comparison between the 42K-wnd (base) and 14K-wnd (current) shown in Fig.~3, we obtain $(1-{\rm corr}) \simeq 0.22$, $(1-R^2) \simeq 0.39$, NRMSE $\simeq 0.62$, and $f_{\rm flip} \simeq 0.29$ for the upper 10\% region, and $(1-{\rm corr}) \simeq 0.30$, $(1-R^2) \simeq 0.51$, NRMSE $\simeq 0.71$, and $f_{\rm flip} \simeq 0.38$ for the upper 50\% region. The systematic increase of all metrics for the larger threshold indicates that the reorganization extends beyond the most intense regions, encompassing a broader spatial area. These results further suggest that, while a weak structural correlation persists, the pattern cannot be captured by a simple linear transformation; instead, it exhibits substantial amplitude changes together with partial sign reversal. This behavior is consistent with a reorganization of the spatial pattern associated with the emergence of ferroelectric-domain structures below $T_{\rm F}$.

\end{document}